\RequirePackage{fix-cm}
\documentclass[runningheads]{llncs}

\usepackage[titletoc,toc,title]{appendix}
\usepackage[usenames,dvipsnames]{color}
\usepackage[usenames,dvipsnames]{xcolor}
\usepackage{alltt}
\usepackage{multicol}
\usepackage{geometry}
\usepackage{float}
\usepackage{multirow}
\usepackage{rotating}
\makeatother
\usepackage{amsfonts}
\usepackage{amssymb}
\usepackage{pgfplots}
\usepackage{tikz}
\usetikzlibrary{arrows,automata}
\usetikzlibrary{arrows}
\usepackage{pgfplotstable}
\usepackage{enumerate}
\usepackage{stmaryrd}
\usepackage{epstopdf}
\usepackage{graphicx}
\usepackage{url}
\usepackage {tikz}
\usetikzlibrary {positioning}
\usepackage{calc}
\usepackage{ifthen}
\usepackage{listings}
\usepackage{mathtools}
\usepackage{syntax}
\usepackage[normalem]{ulem}
\usepackage[normalem]{ulem}
\usepackage{tikz}
\usetikzlibrary{decorations.pathmorphing}
\usetikzlibrary{arrows,shapes}
\usetikzlibrary{shadows,fadings}

\tikzset{snake it/.style={decorate, decoration=snake}}
\usepackage{hyperref}

\def\Msg{\mathit{Msg}}

\def\MessageTransition{$\xRightarrow[]{}_m$}
\def\StatementTransition{$\xRightarrow[]{}_s$}
\def\NetworkTransition{$\xRightarrow[]{}_n$}
\def\NonurgentTransition{$\xRightarrow[]{}_N$}

\def\MessageServers{\it msgsrvs}
\def\Modes{\it modes}

\def\MessageName{m}

\def\ModeName{\mathfrak{m}}
\def\ModeIden{\mathfrak{M}}

\def\WireCon{Wire}
\def\CANCon{CAN}
\lstdefinelanguage{Hrebeca}{
	morekeywords={softwareclass, physicalclass,changemode, mode,inv,msgsrv,guard,delay,statevars, int,float,real,knownrebecs,if,else,self,main,Wire,CAN},
	otherkeywords={=>,<-,<\%,<:,>:,\#,@},
	sensitive=true,
	morecomment=[l]{//},
	morecomment=[n]{/*}{*/},
	morestring=[b]",
	morestring=[b]',
	morestring=[b]"""
}
\begin{document}
	
\title{Hybrid Rebeca: Modeling and Analyzing of Cyber-Physical Systems %\thanks{Supported by organization x.}
	}
\author{Iman Jahandideh\inst{1}%\orcidID{0000-1111-2222-3333}
	 \and
	Fatemeh Ghassemi\inst{1}
	%\orcidID{1111-2222-3333-4444} \and
	Marjan Sirjani\inst{2,3}
	%\orcidID{2222--3333-4444-5555}
	}
\institute{School of Electrical and Computer Engineering, University of Tehran \email{\{jahandideh.iman,fghassemi\}@ut.ac.ir}\\ \and
	School of Innovation, Design and Engineering, M\"{a}lardalen University, V\"{a}ster{\aa}s, Sweden 
	\email{marjan.sirjani@mdh.se}\\
 \and
	 School of Computer Science, Reykjavik University, Reykjavik, Iceland \\
	}

\maketitle

\begin{abstract}
In cyber-physical systems like automotive systems, there are components like sensors, actuators, and controllers that communicate asynchronously with each other.	The computational model of actor supports modeling distributed asynchronously communicating systems. We propose Hybrid Rebeca language to support modeling of cyber-physical systems. Hybrid Rebeca is an extension of actor-based language Rebeca. In this extension, physical actors are introduced as new computational entities to encapsulate physical behaviors. To support various means of communication among the entities, the network is  explicitly modeled as a separate entity from actors. We derive hybrid automata as the basis for analysis of Hybrid Rebeca models. We demonstrate the applicability of our approach through a case study in the domain of automotive systems. We use SpaceEx framework for the analysis of the case study. 
	

\end{abstract}

\section{Introduction}

Embedded systems consist of microprocessors which control a  physical behavior. Ninety-eight percent of all microprocessors are manufactured as components of embedded systems \cite{Embeded}. In such \emph{hybrid} systems, physical and cyber behaviors, characterized as continuous and discrete respectively, affect each other. Cyber-physical sytems (CPSs) are heterogeneous systems with tight interactions between physical and software processes where components in the system usually communicate through network. These systems are used in wide variety of safety-critical applications, from automotive and avionic systems to robotic surgery and smart grids. This makes verifying and analyzing CPSs one of the main concerns while developing such systems. 

Model-based design is an effective technique for developing correct CPSs \cite{derler2012modeling}. It relies on models specifying the behavior of the system often in a formal way. Using models instead of physical realizations of the system, beside reducing the costs of the development, can provide new insights early in the design process and enable analyzing the system behavior in many complex situations that can not easily be reproduced in its real environment. %physical form. 
Furthermore formal and extensive analysis of the model can provide more confidence in the correctness of the system.
%
%Designing CPSs involves trans-disciplinary approaches and is the intersection between physical and cyber entities. 
The heterogeneity of CPSs creates new modeling challenges that stem from interactions between different kinds of components. New theories and tools are needed to facilitate designing and analyzing CPSs. Furthermore, for dealing with such systems with complicated and heterogeneous components, besides expressiveness power, a level of \emph{friendliness} is appealing in design tools. This friendliness can be as important as expressiveness \cite{MarjanFriend}. Friendliness is evaluated by its \emph{faithfulness} to the system it is modeling, and \emph{usability} to the modeler.

Existing modeling frameworks for hybrid systems such as hybrid automata \cite{alur1995algorithmic,henzinger2000theory} and hybrid Petri nets 
\cite{DavidA01} can be used to model CPSs.  %These two models are characterized by their analysis and modeling power, respectively. 
The former has higher analysis power while the latter can be more easily used for %has more modeling power due to its %the highest expressiveness while others provide a 
%higher level of abstraction in 
modeling event-based systems \cite{DavidA01}. 
%A higher level of abstraction, not only increases the usability of a language but also may increase the analyzability of models. 
%As the complexity of a system increases, the size of the model increases exponentially~\cite{Song} which makes its analysis difficult. Furthuremore, 
Due to the existence of network in CPSs, the provided modeling power in these frameworks %level of abstraction 
is not satisfactory for systems composed of many interacting heterogeneous entities. In the domain of automotive, ECUs, sensors and actuators may be connected directly by wire or through a communication media such as a serial bus. Improving the level of abstraction is beneficial to reduce errors introduced during the design process and improve perception of the model. %make perception more easily. 

% TODO:‌ Are actor citations correct?]
The computation model of actors provides a suitable level of abstraction to faithfully model distributed asynchronously communicating systems \cite{agha1985actors,hewitt1972description}. Actors are units of computation which can only communicate by asynchronous message passing. Each actor is equipped by a mailbox in which the received messages are buffered. 
% Application of the actor model as the basis for modeling CPSs, has been followed in Ptolemy \cite{ptolemaeus2014system} and development tools such as \cite{CicirelliNS18} which uses simulation for analyzing the systems. %, and 
%To take advantage of hybrid automata in formal analysis of CPSs, 
%To take advantage of its abstraction, 
We extend the actor-based language Rebeca, with physical behavior to support hybrid systems. %Such an extension is not straightforward. %Realizing such functional requirements while the usability and analyzability of the models are preserved.
%The considered functional requirements for our extension are 1) the value of real variables should be always accessible to their encapsulating actor, 2) a software actor, controlling a continuous behavior, can start, stop or modify the continuous behavior at any time.  
%We discuss that extending actors with real-valued variables or continuous behaviors to realize such functional requirements may complicate the usability and analyzability of a model. 
Additionally, we need to support various types of communication, namely wired connections with no delay, serial buses with deterministic behavior, and wireless communication among the actors. So, we decided not to model the behavior of the network as an actor
within a model, and instead model it 
as a separate entity.

To implement the extended actor model, we propose \emph{Hybrid Rebeca}, as an extension of (Timed) Rebeca \cite{sirjani2004modeling,aceto2011modelling}. Rebeca provides an operational interpretation of actor model through a Java-like syntax. % while  bridging the gap between formal methods and software engineering. 
Its timed extension supports modeling of the computation time, and network delay for message communication. Hybrid Rebeca, extends Timed Rebeca with continuous behaviors based on our extended actor model. 

Hybrid Rebeca defines two types of classes, software and physical. Software classes are similar to reactive classes in Rebeca language where the computational behaviors are defined by message servers. Physical classes in addition to message servers, can also contain different modes, where the continuous behaviors are specified. A physical actor (which is instantiated from a physical class) must always have one active mode. This active mode defines the runtime continuous behavior of the actor. By changing the active mode of a physical actor, it's possible to change the continuous behavior of the actor. In this version, CAN network is defined as network model for communications of the actors. Actors can communicate with each other either through the CAN network or directly by wire. Since CAN is a priority-based network, a priority must be assigned for the messages that are sent through CAN. Real-valued variables are added on which % Three data types \textit{int}, \textit{real}, and \textit{float} are defined in Hybrid Rebeca. The 
continuous behaviors are defined. The modes of physical classes are similar to the concept of locations in hybrid automata, and to solve these behaviors, the semantics of Hybird Rebeca is defined as a hybrid automaton, for which many verification algorithms and tools are available.

The main contribution of the paper can be summarized as providing an actor-based  formalism that supports ``friendliness'' with small number of primitive concepts.  In particular it distinguishes between software and physical actors and supports two types of connections among actors (in principle one could have more types). A tool which automatically derives a hybrid automaton from a given model is implemented, which is suitable for formal reachability analysis. The rest of the paper is structured as follows. The next section defines hybrid automata, actor model and Rebeca language. Section \ref{sec::CPSActorModel} presents our extended actor model for modeling CPSs. In section \ref{Sec:HybridRebeca} the syntax and semantics of Hybrid Rebeca language is defined. Section \ref{sec::CaseStudy} presents our case study and its results. In section \ref{sec::related} we briefly mention some related works. In section \ref{Sec:Discussions} we discuss one of our design decisions for our extend actor model. The conclusion is presented in section \ref{Sec:Concolusion}.
  
%TODO::complete
%We evaluate the effectiveness of our approach by comparing our modeling framework to hybrid automata. To this aim, we consider modularity, as a metric of usability, and analyzability. We show that by using our framework, the cost of improving and modifying models is vastly reduced compared to modeling in hybrid automata. We also show that modeling high level concepts of Hybrid Rebeca like message passing and message buffering directly in hybrid automata can hugely decrease the analyzability of the models. %due to increase in dimensions of the model.  
%Furthermore, the abstraction resulted from choosing actors as the basic units of computation, offers more friendliness towards cyber-physical systems compared to the low-level languages like hybrid automata.

%TODO::complete
%Our evaluation results indicate that   
%\section{Background/Motivation}
%\subsection{Motivation}

\section{Preliminaries}\label{Sec:Preliminaries}
As we define the semantics of our framework based on hybrid automata, we first provide an overview on this model and then explain actor model and Timed Rebeca.

\subsection{Hybrid Automata}

\def\HAExampleTemp{\textit{t}}
\def\HAExampleOff{\textit{off}}
\def\HAExampleOn{\textit{on}}

\def\HAExampleOnToOffGuard{$t == 22$}
\def\HAExampleOnInvariant{$t \le 22$}

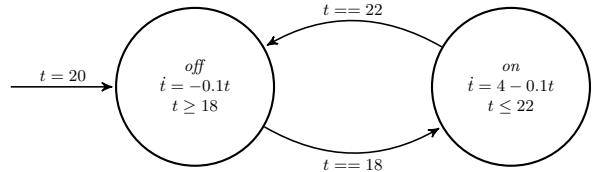
\begin{figure}[h]
	\centering
	
\usetikzlibrary{arrows,shapes,automata,petri,positioning,calc}
\usetikzlibrary{shapes.multipart}

\def\Tail{\textit{tail}}
\def\Head{\textit{head}}

\newcommand{\Element}[1]{\textit{elm{#1}}}

\def\Msg{\textit{msg}}
\def\Buffer{\textit{buf}}
\def\BufferFull{\textit{bufful}}
\def\Busy{\textit{busy}}
\def\EmptyBuffer{\BufferFull==0}
\def\Size{\textit{size}}
\def\SyncOne{\textit{sync1}}
\def\SyncTwo{\textit{sync2}}
\def\And{\&\&}
\tikzset{
    place/.style={
        circle,
        thick,
        draw=white,
        fill=white,
        minimum size=6mm,
    },
        state/.style={
        circle,
        thick,
        draw=black,
        fill=white,
        minimum size=3cm,
    },
    invisibleState/.style={
        circle,
        thick,
        draw=white,
        fill=white,
        minimum size=6mm,
    },
}

\begin{tikzpicture}[->,>=stealth',shorten >=1pt,auto,node distance=6cm,semithick,scale=.7, transform shape]
  \tikzstyle{every state}=[fill=red,draw=none,text=white]

\node[state] (On)                  { \begin{tabular}{c} \textit{on} \\ $\dot{t}=4-0.1t$ \\ $t \le 22$ \end{tabular} };
\node[state] (Off)           [left of = On]   { \begin{tabular}{c} \textit{off} \\ $\dot{t}=-0.1t$ \\ $t\ge 18$ \end{tabular} };
\node[invisibleState] (Initial)           [left of = Off,left=-2.5 ]   {};

\path (On) edge [bend right, above]  node{$t==22$} (Off);
\path (Off) edge [bend right, below] node {$t == 18$} (On);
\path (Initial) edge [above] node {$t = 20$} (Off);

%   \path (L0) edge              node {0,1,L} (B)
%             edge              node {1,1,R} (C)
%         (B) edge [loop above] node {1,1,L} (B)
%             edge              node {0,1,L} (C)
%         (C) edge              node {0,1,L} (D)
%             edge [bend left]  node {1,0,R} (E)
%         (D) edge [loop below] node {1,1,R} (D)
%             edge              node {0,1,R} (L0)
%         (E) edge [bend left]  node {1,0,R} (L0);
\end{tikzpicture}
	\caption{A hybrid automaton for a heater which consists of two locations (modes) named \HAExampleOn\ and \HAExampleOff. Each location defines a flow and an invariant on the variable \HAExampleTemp\ which is the temperature. The mode of the heater changes by means of guarded transitions between the locations. The initial location is \HAExampleOff\ and the initial value of \HAExampleTemp\ is $20$.}
	\label{fig::HATExample}
\end{figure}

\emph{Hybrid automata} (HA) \cite{alur1995algorithmic,henzinger2000theory} is a formal model for systems with discrete and continuous behaviors. Informally a hybrid automaton is a finite state machine consisting of a set of real-valued variables, modes and transitions. Each mode, which we also call \emph{location}, defines a continuous behavior on the variables of the model. The continuous behaviors or \emph{flows} are usually described by ordinary differential equations which define how the values of the variables change with time. Transitions act as discrete actions between continuous behaviors of the system, where the variables can change instantaneously. In Fig.~ \ref{fig::HATExample} a hybrid automaton for a simple heater model is presented. The variable \HAExampleTemp\ represents the temperature of the environment. The locations named \HAExampleOff\ and \HAExampleOn\ define the continuous behavior of the temperature when the heater is off and on, respectively. For each location, the flow of the temperature is defined accordingly. The transition with the guard \HAExampleOnToOffGuard\ states that when the temperature is equal to $22$ the heater \emph{can} be turned off. In hybrid automaton, the choice between staying in one location and taking an enabled transition is nondeterministic. To make the turning off behavior deterministic, the invariant \HAExampleOnInvariant\ is defined in the \HAExampleOn\ location. This invariant states that the heater can only stay in this location as long as the temperature is less than $22$. The turning on behavior of the heater is defined similarly. Initially the heater is off and the temperature is $20$.

Let a valuation $v: V\rightarrow {\it Val}$ be a function that assigns a value to each variable of $V$ where $\it Val$ is the set of values, defined by the context. We denote the set of valuations on the set of variables $V$ as $\mathcal{V}(V)$. Formally a hybrid automaton is defined by the tuple $({\it Loc},{\it Var},{\it Lab}, \Rightarrow, {\it Flws}, {\it Inv}, {\it Init})$ as follows:

\def\HALoc{{\it Loc}}
\def\HAVar{{\it Var}}
\def\HALab{{\it Lab}}
\def\HAFlws{{\it Flws}}
\def\HAInv{{\it Inv}}
\def\HAInit{{\it Init}}

\begin{itemize}
	\item  $\it Loc$ is a finite set of locations,
	\item $\it Var$ is a finite set of real-valued variables, %. We use $\it Val$ as the set of valuations for these variables,
	\item $\it Lab$ is a finite set of synchronization labels.
	\item $\Rightarrow$ is a finite set of transitions. A transition is a tuple $(l,a,\mu,l')\in {\Rightarrow}$ where $l\in {\it Loc}$ is the source location, $l'\in Loc$ is the destination location, $a\in Lab$ is a synchronization label and $\mu \in \mathcal{V}({\it Var})^{2}$ is a transition relation on variables. The elements of $\mu = (v,v')$ represents the valuation of the variables before and after taking the transition. In some models, like in our example, this transition relation is represented with a guard and a set of assignments on the variables. The guard defines the valuation $v$ and the assignments define the valuation $v'$. 
	\item $\it Flws$ is a labeling function that assigns a set of flows to each location $l\in Loc$. Each flow is a function from $\mathbb{R}^{\ge 0} \rightarrow \mathcal{V}({\it Var})$. Each flow specifies how the values of variables evolve over time. A flow is usually defined by a doted variable $\dot{v}$ which represents the first derivative.
	\item $\it Inv$ is a labeling function that assigns an invariant ${\it Inv}(l)\subseteq \mathcal{V}({\it Var})$ to each location $l\in{\it Loc}$,
	\item $\it Init$ is a labeling function that assigns an initial condition ${\it Init}(l) \subseteq \mathcal{V}({\it Var})$  to each location $l\in{\it Loc}$. 
\end{itemize}

 In the example given in Fig.~\ref{fig::HATExample}, the locations and the variables are defined as $ \HALoc = \{\HAExampleOn, \HAExampleOff\}$ and $\HAVar = \{ \HAExampleTemp \}$ respectively. Since our example only consists of a single automaton, $\HALab = \emptyset$ and the labels over the transitions are $\epsilon$ which is not shown for brevity. Also the transition relations only consist of guards and the assignments are empty. The flows and the invariant of each location is defined on the location itself. The initial condition for location \HAExampleOff\ is $\HAInit(\HAExampleOff)=\{t=20\}$ and for location \HAExampleOn\ is $\HAInit(\HAExampleOff)=\emptyset$. Note that in our language, we do not use primed variables of the form $v'$ to represent valuation after discrete transitions. We use $v'$ instead of $\dot{v}$ to represent the first derivative of variable $v$
%{\color{red}		
%\begin{definition}[Urgent location and urgent transition] \label{def:UrgentDef}
%	We define an \emph{urgent location} in hybrid automata as a location where no delay is allowed, inspired by \cite{behrmann2004tutorial}. Upon entering an urgent location, there must be at least one enabled transition and immediately one of these enabled transitions must be taken. We also define an \emph{urgent transition} as a transition that must be taken immediately upon entering its source location. A location that has at least one outgoing urgent transition, is also considered an urgent location. Hybrid automata do not have explicit notions for urgent location and urgent transition, but with appropriate invariants, flows and guards these behaviors can be achieved.
%\end{definition}
% }

\subsection{Actor Model and Timed Rebeca}
Actor model is used for modeling distributed systems. It was originally proposed by Hewitt \cite{hewitt1972description}. In this model actors are self-contained and concurrent \cite{agha1985actors} and can be regarded as units of computation. Any communication is done through asynchronous message passing on a fair medium where message delivery is guaranteed but is not in-order. This model abstracts away the network effects like delays, message conflicts, node crashes, etc. In this model each actor has a an address and a mailbox which stores the received messages. The behavior of an actor is defined in its message handlers, called \emph{methods}. The methods are executed by processing the messages.
%Actor services these messages in a FIFO manner.

To extend the actor model with hybrid concepts for specifying CPSs, %. To illustrate our actor model in application, 
we use Rebeca as our basis framework and hence, use the terms actor model and Rebeca interchangeably in this paper. Rebeca \cite{sirjani2004modeling} is a formal actor-based modeling language and is supported by model checking tools to bridge the gap between formal methods and software engineering. Rebeca provides an operational interpretation of actor model through a Java-like syntax. It also offers a compositional verification approach to deal with the state-space explosion problem in model checking. Because of its design principle it is possible to extend the core language for a specific domain \cite{sirjani2011ten}. For example, different extensions have been introduced in various domains such as probabilistic systems \cite{varshosaz2012modeling}, real-time systems \cite{aceto2011modelling}, software product lines \cite{Sabouri}, and broadcasting environment \cite{bRebeca}. 

In Rebeca, actors are called rebecs and are instances of \emph{reactive classes} defined in the model. Rebecs communicate with each other through asynchronous message passing and a Rebec's mailbox is modeled by a message queue. A reactive class consists of \emph{known rebecs} to specify the rebecs it can communicate with, \emph{state variables} to maintain the internal state, and \emph{message servers} to define the reaction of the rebec on the received messages. The computation in a rebec takes place by removing a message from the message queue and executing its corresponding message server.

Timed Rebeca \cite{aceto2011modelling} is an extension of Rebeca for distributed and asynchronous systems with timing constraints. It adds the timing concepts \textit{computation time}, \textit{message delivery time} and \textit{message expiration}. These concepts are materialized by new constructs: \textit{delay}, \textit{after}, and \textit{deadline}.
In Timed Rebeca model, each rebec has it's own local clock which can be considered as synchronized distributed clocks.
The \textit{delay} statement models the passage of time during the execution of a message server. Statements \textit{after} and \textit{deadline} are used in conjunction with send statements and specify the network delay and the message deadline, respectively.

\def\HeatingCtrl{\textit{HeatingCtrl}}
\def\HeatingCtrlChangeTemp{\textit{changeTemp}}
\def\HeatingCtrlOn{\textit{setTemp}}
\def\HeatingCtrlOff{\textit{turnOff}}
\def\HeatingCtrlInitial{\textit{initial}}
\def\HeatingCtrlAway{\textit{away}}
\def\HeatingCtrlTemp{\textit{temp}}
\def\HeatingCtrlCoeff{\textit{coeff}}
\def\HeatingCtrlStatus{\textit{status}}
\def\HeatingCtrlTargetTemp{\textit{targetTemp}}

\section{Actor Model for CPSs}  \label{sec::CPSActorModel}
%To specify a CPS, there 

Extending actor model for modeling cyber-physical systems can be divided to two parts, \textit{offering more concrete models for network}, and \textit{extending actors with physical behaviors}.

Rebeca offers a fair and nondeterministic network model. For many application of CPSs this network model is too abstract or completely invalid. For example Control Area Network (CAN) \cite{pfeiffer2008embedded} protocol is a dominant networking protocol in automotive industry, which can not be faithfully modeled by Rebeca's network model as by this protocol, messages are deterministically delivered to their receivers. Modeling the network as an explicit actor, does not guarantee determinacy of message deliveries as the network actor is executed concurrently with other actors,  therefore its determinacy is affected by the interleaving semantics. In other words, the massage delivery sequence by the network actor depends on the execution order of sending actors. So we modeled the network as a separate entity from the actors.

To extend the actor model with physical behaviors, we decided to separate physical actors from software actors. In this approach software actors will be similar to Timed Rebeca actors and the physical behaviors are defined in separate physical actors. Physical actors are similar to a hybrid automaton in syntax and semantics. Like a hybrid automaton, each physical actor consists of a set of modes. Each mode defines its flows, invariant, guard and actions where actions are a set of statements. The actions  are the effect of the mode, when the continuous behavior is finished. 
A physical actor can only be in one mode (characterizing a specific continuous behavior) at any moment. 
In this approach the physical behavior of a system can easily be started, stopped or changed by changing the active modes of physical actors, either by the actor itself or by a request from another actor.

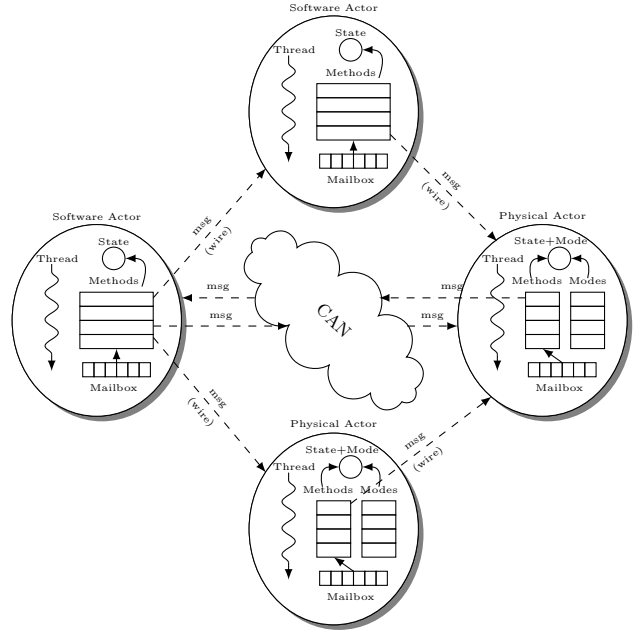
\begin{figure}[tbp]
	\centering
	\usetikzlibrary{decorations.pathmorphing}
\usetikzlibrary{arrows}
\usetikzlibrary{shadows,fadings}
\usetikzlibrary{calc}
\usetikzlibrary{matrix}
%\usetikzlibrary{math}

\tikzset{snake it/.style={decorate, decoration=snake}}
\def\QueueXOffset{0.2}
\def\MQSize{0.2,-0.25}

\newcommand{\DrawVBoxes}[4]
{              
	\draw (#1, #2) rectangle + (#4);
	\draw ( #1 + 1* #3, #2) rectangle + (#4);
	\draw ( #1 + 2* #3, #2) rectangle + (#4);
	\draw ( #1 + 3* #3, #2) rectangle + (#4);
	\draw ( #1 + 4* #3, #2) rectangle + (#4);
	\draw ( #1 + 5* #3, #2) rectangle + (#4);
	\node at (#1 + 0.55, #2 - 0.45) {\tiny Mailbox};
}

\newcommand{\DrawHBoxes}[4]
{      
	\draw (#1, #2) rectangle + (#4);
	\draw ( #1 , #2 + 1* #3) rectangle + (#4);
	\draw ( #1 , #2 + 2* #3) rectangle + (#4);
	\draw ( #1 , #2 + 3* #3) rectangle + (#4);
}

\newcommand{\DrawSoftwareActor}[2]
{      
	\node at (#1+1,#2+3.85) {\tiny Software Actor};
	\draw[drop shadow={fill=black}]  (#1+1,#2+2) ellipse (1.5 and 1.7);
	\draw[fill=white]  (#1+1,#2+2) ellipse (1.5 and 1.7);
	 \draw[snake it] (#1+0.2,#2+1.4) -- (#1+0.2,#2+3);
	\draw[-latex] (#1+0.2,#2+1.4) -- (#1+0.2,#2+1.1);
	\node at (#1+0.3,#2+3.1) {\tiny Thread};
	%methods
	\DrawHBoxes{#1+0.7}{#2+1.5}{0.25}{1.3,0.25}
	\node at (#1+1.3,#2+2.7) {\tiny Methods};
	
	%queue regular
	\DrawVBoxes{#1+0.75}{#2+1.25}{\QueueXOffset}{\MQSize}
	
	\draw[-latex] (#1+1.35,#2+1.25)--(#1+1.35,#2+1.5);
	%state
	\draw  (#1+1.3,#2+3.1) ellipse (0.2 and 0.2);
	\node at (#1+1.3,#2+3.4) {\tiny State};
	\draw[-latex] (#1+1.8,#2+2.6) .. controls (#1+1.9,#2+2.9) and (#1+1.9,#2+3.1)  .. (#1+1.5,#2+3.1);
}

\newcommand{\DrawPhysicalActor}[2]
{      
	\node at (#1+1,#2+3.85) {\tiny Physical Actor};
	\draw[drop shadow={fill=black}]  (#1+1,#2+2) ellipse (1.5 and 1.7);
	\draw[fill=white]  (#1+1,#2+2) ellipse (1.5 and 1.7);
	 \draw[snake it] (#1+0.2,#2+1.4) -- (#1+0.2,#2+3);
	\draw[-latex] (#1+0.2,#2+1.4) -- (#1+0.2,#2+1.1);
	\node at (#1+0.3,#2+3.1) {\tiny Thread};
	%methods
	\DrawHBoxes{#1+0.7}{#2+1.5}{0.25}{0.6,0.25}
	\DrawHBoxes{#1+1.5}{#2+1.5}{0.25}{0.6,0.25}
	\node at (#1+0.9,#2+2.7) {\tiny Methods};
	\node at (#1+1.8,#2+2.7) {\tiny Modes};
	%queue
	\DrawVBoxes{#1+0.75}{#2+1.25}{\QueueXOffset}{\MQSize}

	\draw[-latex] (#1+1.35,#2+1.25)--(#1+1,#2+1.5);
	%state
	\draw  (#1+1.3,#2+3.1) ellipse (0.2 and 0.2);
	\node at (#1+1.15,#2+3.4) {\tiny State+Mode};
	\draw[-latex] (#1+1.8,#2+2.75) .. controls (#1+1.8,#2+2.75) and (#1+1.9,#2+3.1)  .. (#1+1.5,#2+3.1);
	\draw[-latex] (#1+0.8,#2+2.75) .. controls (#1+0.8,#2+2.75) and (#1+0.7,#2+3.1)  .. (#1+1.1,#2+3.1);
}

\newcommand{\DrawMessageSend}[4]
{
	%\draw[-latex,dashed] (#1,#2) --(#3,#4);
	%\draw  ( 0.5 * ((#2) - (#1)) ,(#3) - (#4)/2 ) rectangle +(\MQSize);
	%\draw  (#5) rectangle +(\MQSize);
	\draw[-latex,dashed] (#1,#2)  -- (#3,#4) node [midway,above, sloped] {\tiny msg};
}

\newcommand{\DrawMessageSendWire}[4]
{
	%\draw[-latex,dashed] (#1,#2) --(#3,#4);
	%\draw  ( 0.5 * ((#2) - (#1)) ,(#3) - (#4)/2 ) rectangle +(\MQSize);
	%\draw  (#5) rectangle +(\MQSize);
	\draw[-latex,dashed] (#1,#2)  -- (#3,#4) node [midway,above, sloped] {\tiny msg} 
	node [midway,below, sloped] {\tiny (wire)} ;
}

\newcommand{\DrawCreate}[4]
{
	\draw[-latex] (#1,#2)  -- (#3,#4) node [midway,above, sloped] {\tiny Create};
}

\usetikzlibrary{patterns}
\begin{tikzpicture}[scale=0.75, transform shape]
%-----------actor 1

\DrawSoftwareActor{-1.7}{0}
\DrawSoftwareActor{2.5}{3.7}
\DrawPhysicalActor{2.5}{-3.7}
\DrawPhysicalActor{6.2}{0}

%effect
%-----
\DrawMessageSendWire{0.3}{2.4}{2.3}{4.7}
\DrawMessageSendWire{0.3}{1.7}{2.3}{-0.7}

\DrawMessageSendWire{3.8}{-1.25}{6.3}{0.65}

\DrawMessageSend{4.8}{1.9}{5.7}{1.9}
\DrawMessageSend{0.3}{1.9}{2.7}{1.9}

\DrawMessageSend{6.9}{2.4}{4.3}{2.4}
\DrawMessageSend{2.1}{2.4}{0.8}{2.4}

\DrawMessageSendWire{4.5}{5.3}{6.4}{3.4}

%\draw[-latex,dashed] (2,2) --(4.6,1.5);
%\draw  (3.1,1.8) rectangle (3.2,2.05);
%\draw[-latex,dashed] (2,2.1) --(4.6,2.5);
%\draw[fill=black!20]   (3.3,2.5) rectangle (3.4,2.75);
%\draw[-latex] (0.5,4.5)  -- (3.5,3.8);
%\draw[] (-1,5)  edge node[above,sloped] {\tiny Create} (3,3.5);
%-----------------------------------
%\draw[-latex] (2.1,0.85)  .. controls (2.3,0.5) and (2.8,1.1)  ..  (2.3,1.2);%
%\node at (2.6,0.7) {\tiny wait};

\node[rotate=-45,cloud, cloud puffs=12, cloud ignores aspect, minimum width=4cm, minimum height=2cm, align=center, draw] (cloud) at (3.5, 2) {CAN};

\end{tikzpicture}
	\caption{Hybrid Rebeca model: each actor has its own thread of control, message queue, and ID. In addition to these, physical actors have modes that are defined by a guard, an invariant, and actions. Actors can communicate with each other either by sending messages through CAN or directly by wire.}\label{Fig::ourActor}
\end{figure}

As we focus on automotive systems, to make the network specification more concrete, in the first step we consider the CAN protocol in our language. CAN is a serial bus network where nodes can send messages at any moment. When multiple nodes request to send a message at the same time, only the message with the highest priority is accepted and sent through the network. After a message is sent, the network chooses another message from the requested messages. The messages are sent through the network one by one. As messages in this protocol must have unique priorities, messages are deterministically communicated. Furthermore, we assume that all CAN nodes implement a priority-based buffer. This simplifies the network model which can be represented by a single global priority-based queue \cite{davis2007controller}. %; messages with a same time-stamp are deterministically inserted into the queue based on their priorities.
To implement this protocol, a unique priority must be assigned to each message and the communication delay between each two communicating actors must be specified. These specifications can be defined outside of the actors so that actors become agnostic about the underlying network of the model. This will also make the model more modular, since it is easier to change the network of the system without modifying the actors. Not all the actors communicate through CAN. Some of the actors may be connected by wire and have direct communication with each other. In our language, both types of communication are considered, and actors can communicate with each other either via wire or CAN. All messages, irrespective of the communication medium, are eventually inserted to their receiver's message queue. If two or more simultaneous messages (from wire or CAN) got inserted into a message queue, the resulting ordering will be nondeterministic. Note that there can not be two simultaneous messages from CAN, since CAN is a serial bus. The resulting hybrid Rebeca model has been illustrated in Fig.~\ref{Fig::ourActor}.

%Not all the actors communicate through CAN. Some of the actors may be connected by wire and have direct communication with each other. In Rebeca, actors have a FIFO message scheduler and execution of messages are directly dependent on their receiving order. For the messages that are sent to one actor by concurrently executing actors, this receiving order is nondeterministic, so the behavior of the actor will also be nondeterministic. To resolve this nondeterminism, modelers can use a priority-based message scheduler instead of the default FIFO scheduler. This scheduler works the same way as the CAN protocol and chooses the highest priority message no matter what the ordering is.  %%%An embedded system is modeled by a set of software and physical actors. As the number of embedded devices in a system maybe dynamic, a software actor can create both software and physical actors. Furthermore, a physical actor can create physical actors to model some physical phenomena, e.g, a falling drop can be divided into several drops upon collision with ground.  Upon processing a message, a software actor may create new actors, send messages, or update its state variables. A physical actor can also create new physical actors, send messages, or update its mode and state variables.   

\section{Hybrid Rebeca}\label{Sec:HybridRebeca}
Hybrid Rebeca is an extension of Timed Rebeca to support physical behaviors. Timed Rebeca supports  deadline specification for messages, but  because of technical issues, Hybrid Rebeca does not support this feature. Other timing behaviors like network delay and computation time are supported. 
%To simplify the semantics of our language, we intentionally do not %resolve the non-determinism introduced by the simultaneous arrival of messages over wire to an actor 
%consider a priority-based message scheduling for actors in this extension.  

%%%As mentioned earlier, in Rebeca's terminology actor are called rebec and through out this section we will use the term rebec for an actor instance. 
In Hybrid Rebeca we have two types of rebecs: software and physical. These rebecs communicate through asynchronous message passing. Each rebec has a queue for messages, and services the message at the head of the queue by executing the corresponding message server for that message.

Software rebecs are for modeling software (discrete) behaviors. These rebecs are reactive and self-contained and they can have multiple message servers. Physical rebecs are for modeling physical (continuous) behaviors and the physical behaviors are defined by their modes. For physical rebecs a reserved message server is defined for changing the rebec's active mode.% These rebecs can't have any user defined message servers.

Hybrid Rebeca has the concept of class for rebecs, and rebecs of the model are instantiated from these classes in the main block. In the instantiation phase the connection type of the rebecs with each other must be defined. For now our language supports only two types of connection: CAN and wire. When rebecs communicated through wire, the communication delay is considered to be zero. After instantiating rebecs, the CAN specification must be defined.% This includes message priorities and network delay for communications. 

\subsection{Syntax}
A Hybrid Rebeca model definition consists of a set of \textit{classes} and a \textit{main} block, where classes define different types of rebecs and \textit{main} specifies the initial configuration and CAN specification.

The syntax of Hybrid Rebeca is presented in Fig.~\ref{Fig::HybridRebecaGrammar}. The syntax of a software class is similar to a reactive class in Timed Rebeca, which resembles a class definition in Java. A software class consists of a set of known rebecs, state variables and message servers. The known rebecs are the rebecs that an instance of this class can send message to. The syntax of  message servers is like a method in object-oriented languages, expect that they have no return value. 

The core statements of our language are variable/mode assignment, conditional, delay, and method calls. An actor can send a message asynchronously to other rebecs through method calls.% are asynchronous message passing .  

\def\Self{$\mathsf{self}$}
\def\None{$\mathsf{none}$}
\def\SetModeMessage{\textit{setMode}}
\def\SetModeStatement{\textit{setmode}}

A physical class is similar to a software class except that it also contains the definition for physical modes. The structure of a $\mathsf{mode}$ resembles a location in hybrid automata, and it consists of invariant, flows, guard and a set of actions. The comparison expression of an invariant and a guard expression are specified by the reserved words $\mathsf{inv}$ and $\mathsf{guard}$, respectively. The actions following the guard expression, which are expressed as a statement block, define the behavior of the rebec upon leaving the corresponding mode. We remark that the next entering mode is either explicitly defined by the user through the statement $\mathsf{\SetModeStatement}$ in the statement block or the default mode \None\ if it was not specified. Mode \None \ is a special mode defined in all physical rebecs. This mode represents an idle behavior and its flows are defined as zero. Activating this mode %as the active mode 
can be interpreted as stopping the physical behavior of a physical rebec. Other rebecs can change the mode of a physical rebecs by sending the message \SetModeMessage. % \ is a special kind of message that is used for changing .  

Three primitive data types are available in Hybrid Rebeca: \textit{int}, \textit{real}, and \textit{float}. Variables of types \textit{int} and \textit{real} are only allowed in software and physical classes, respectively. Message parameters and state variables can be only defined of primitive type with some restrictions. Variables of type \textit{float} can be used in both types of classes. Mathematically the float and real values are the same. However to each real variable, a flow is assigned which determines how its value evolve with time. A float variable can be used to capture the value of a real variable in different snapshots. This can be used in communication with software rebecs. The value of a float variable can be changed only by assignment, but the value of a real variable can be changed by both assignment and the flow defined on the variable. The assignment of a real value to float is managed implicitly in the semantics and no explicit casting is needed. %However the callee message server should be parametrized by a float variable.

Every class definition must have at least one message server, named \textit{initial}. In the initial state of the model, an \textit{initial} message is put in all rebecs's message queue. The state variables and behavior of rebecs are initialized by executing this message server. The keyword \Self \ is used for sending a message to the rebec itself. 

%CHECK ::well-formeness

Rebecs are instantiated in the \textit{main} block of the model. To instantiate a rebec, its known rebecs must be specified to be binded to the appropriate instances. Furthermore for each known rebec, the connection type must also be specified, which can either be $\it CAN$ or $\it Wire$. For example, by the statement ${\it A}~~a~~(@{\it Wire}~b,~~ @{\it CAN}~c):()$, a rebec named $a$ is instantiated from the class $A$ that its known rebecs are $b$ and $c$ while the communication from $a$ to $b$ is through wire and $a$ to $c$ through CAN. We remark that the connection type between two rebecs can be different for each communication direction. The pair of parenthesizes $()$ after the colon represents the parameters of the \textit{initial} message server (which is empty in this case).  After instantiation, the \textit{CAN} specification is defined on the messages that may be transmitted through CAN. This specification consists of two parts. First the \emph{priorities} of these messages must be specified. To this aim, a unique priority is assigned to a message. For example the statement $a~~b.m~~1;$ means that a message sent from rebec $a$ to rebec $b$ containing the message server name $m$ has a priority of $1$.  A lesser number indicates a higher priority. After the priorities, the network \emph{delays} of CAN communications are specified. For instance the statement $a~~b.m~0.01;$ expresses that the communication delay of sending a message from $a$ to $b$ containing the message server name $m$ is $0.01$.% The values of delays are relative and are interpreted in terms of the system model. %Intuitively, depending of the length of a message, modeled by its parameters, the delay varies. 

\def\HighlightColor{green}
\begin{figure}[H]
\begin{center}
\fbox{\parbox{\columnwidth-5mm}{\vspace{-2mm}
% \begin{lstlisting}[captionpos=b,frame = single,  mathescape,lineskip=.05cm,tabsize=2, language=rebeca]
{\footnotesize
\begin{align*}
	\mathrm{Model} &\Coloneqq~(\mathrm{SClass}~|~\textcolor{\HighlightColor}{\mathrm{PClass}})^+~\mathrm{Main}~\\
	\mathrm{Main} &\Coloneqq ~\mathsf{main}~ \{\mathrm{InstanceDcl}^*~ \textcolor{\HighlightColor}{\mathrm{CANSpec}}\} \\
	\mathrm{InstanceDcl} &\Coloneqq ~\mathrm{C}~ \mathrm{r}~(\mathrm{\langle \textcolor{\HighlightColor}{~\mathsf{\text{@}CAN}~|~\mathsf{\text{@}Wire}~}~ r\rangle}^* )~\colon(\mathrm{\langle c\rangle}^*)\\
	\mathrm{CANSpec} &\Coloneqq~~\textcolor{\HighlightColor}{\mathsf{CAN}~\{\mathrm{Priorities}~\mathrm{Delays}\}}\\
	\mathrm{Priorities} &\Coloneqq~~\textcolor{\HighlightColor}{\mathsf{priorities}~\{ \langle{\mathsf{r~r.\MessageName~c;} }\rangle^*\}}\\
	\mathrm{Delays} &\Coloneqq~~\textcolor{\HighlightColor}{\mathsf{delays}~\{ \langle{\mathsf{r~r.\MessageName~c;} }\rangle^*\}}\\
	%\mathrm{Class} &\Coloneqq~~\mathrm{SClass}~|~\textcolor{\HighlightColor}{\mathrm{PClass}} \\
	\mathrm{SClass} &\Coloneqq~ \textcolor{\HighlightColor}{\mathsf{softwareclass}} ~\mathrm{C}~ \{\mathsf{KnownRebecs}~ \mathsf{Vars}~ \mathsf{MsgSrv}^* \}\\
	\mathrm{PClass} &\Coloneqq~ \textcolor{\HighlightColor}{\mathsf{physicalclass}} ~\mathrm{C}~ \{\mathsf{KnownRebecs}~ \mathsf{Vars}~ \mathsf{MsgSrv}^*~\textcolor{\HighlightColor}{\mathsf{Mode}^*} \}\\
	\mathrm{KnowRebecs} &\Coloneqq~ \mathsf{knownrebecs}~\{\mathrm{VarDcl}^* \}\\
	\mathrm{Vars} &\Coloneqq \mathsf{statevars}~ \{\mathrm{VarDcl}^* \}\\
	\mathrm{VarDcl} &\Coloneqq \mathrm{T}~ \langle\mathrm{v}\rangle^+ ;\\
	\mathrm{MesgSrv} &\Coloneqq \mathsf{msgsrv} ~\mathrm{\MessageName}~(\mathrm{\langle T~v\rangle}^*)~\{\mathrm{Stmt}^* \}\\
	\mathrm{Mode} &\Coloneqq~ \textcolor{\HighlightColor}{\mathsf{mode}~ \mathrm{\ModeName}~\{\textcolor{\HighlightColor}{\mathsf{inv}(\mathrm{e})}~(\textcolor{\HighlightColor}{v'=e})^+~\textcolor{\HighlightColor}{~\mathsf{guard} (\mathrm{e})}}~\mathrm{MSt} \}\\
	%\mathrm{Invariant}~&\Coloneqq~\textcolor{\HighlightColor}{\mathsf{inv}(\mathrm{e})}\\
	%\mathrm{Flow} &\Coloneqq \textcolor{\HighlightColor}{v'=e}\\
	%\mathrm{Guard} &\Coloneqq \textcolor{\HighlightColor}{~\mathsf{guard} (\mathrm{e})}\\
	\mathrm{Stmt}
	&\Coloneqq~
	\mathrm{v}=\mathrm{e};~|~\mathrm{Call};~|~\mathsf{if}(\mathrm{e})~\mathrm{MSt}~[\mathsf{else}~\mathrm{MSt}]~|~ \mathsf{delay}(\mathrm{t});~|~ \textcolor{\HighlightColor}{\mathsf{setmode}(\mathrm{\ModeName});}\\
	\mathrm{Call}
	&\Coloneqq~\mathsf{r.\MessageName(\langle e\rangle ^*)}~|~~\textcolor{\HighlightColor}{\mathsf{r.\SetModeMessage(\mathrm{\ModeName})}}\\
	\mathrm{MSt}
	&\Coloneqq~\{\mathrm{Stmt}^*\}~|~ \mathrm{Stmt} 
\end{align*}
}}}
\end{center}
\vspace{-2mm} \caption{Abstract syntax of Hybrid Rebeca. The main differences in syntax compared to Timed Rebeca, are highlighted with color \textcolor{\HighlightColor}{\HighlightColor}. Angle brackets $\langle~\rangle$ denotes meta parenthesis, superscripts $+$ and $*$ respectively are used for repetition of one or more and repetition of zero or more times. Combination of  $\langle~\rangle$ with repetition is used for comma separated list. Brackets $[~]$ are used for optional syntax. Identifiers $C$, $T$, $\MessageName$, $\ModeName$, $v$, $c$, $r$ and $e$ respectively denote class, primitive type, method name, mode name, variable, constant, and rebec name, respectively; and $e$ denotes an expression.
\label{Fig::HybridRebecaGrammar}}
\end{figure}

\def\WheelClass{\textit{Wheel}}
\def\WheelClassTorque{\textit{trq}}
\def\WheelClassSetTorque{\textit{setTrq}}
\def\WheelClassController{\textit{ctlr}}
\def\WheelClassSPD{\textit{spd}}
\def\WheelClassTime{\textit{t}}
\def\WheelClassBrakeMode{\textit{Rolling}}
\def\WheelClassNoBrakeMode{\textit{NoBrake}}
\def\WheelControllerClass{\textit{WCtlr}}
\def\WheelControllerClassWheel{\textit{w}}
\def\WheelControllerClassGlobalController{\textit{bctrl}}
\def\WheelControllerClassSPD{\textit{wspd}}
\def\WheelControllerClassID{\textit{id}}
\def\WheelControllerClassSetWheelSPD{\textit{setWspd}}
\def\WheelControllerClassApplyTorque{\textit{applyTrq}}
\def\WheelControllerClassSliprate{\textit{slprt}}
\def\WheelControllerClassRequestedTorque{\textit{reqTrq}}
\def\WheelControllerClassVehicleSpeed{\textit{vspd}}
\def\WheelControllerClassWRAD{\textit{WRAD}}
\def\BrakePedalClass{\textit{Brake}}
\def\BrakePedalBraking{\textit{Braking}}
\def\BrakePedalBrakePercent{\textit{bprcnt}}
\def\BrakePedalMaxBrakePercent{\textit{mxprcnt}}
\def\BrakePedalTimer{\textit{t}}
\def\BrakePedalRate{\textit{r}}
\def\BrakePedalGlobalController{\textit{bctrl}}
\def\GlobalBrakeControllerClass{\textit{BrakeCtrl}}
\def\GlobalBrakeControllerSetSpeed{\textit{setWspd}}
\def\GlobalBrakeControllerSetBrakePercent{\textit{setBprcnt}}
\def\GlobalBrakeControllerWCTLRR{\textit{wctlrR}}
\def\GlobalBrakeControllerWCTLRL{\textit{wctlrL}}
\def\GlobalBrakeControl{\textit{control}}
\def\GlobalBrakeBrakePercent{\textit{bprcnt}}
\def\ClockClass{\textit{Clock}}
\def\WheelRRebec{\textit{wR}}
\def\WheelLRebec{\textit{wL}}
\def\WheelControllerRRebec{\textit{wctlrR}}
\def\WheelControllerLRebec{\textit{wctlrL}}
\def\BrakeControllerRebec{\textit{bctlr}}
\def\BrakeRebec{\textit{brake}}
\def\ClockRebec{\textit{clock}}
\def\AtWire{\textit{@Wire}}
\def\AtCAN{\textit{@CAN}}

\subsubsection{Example}\label{subsec::example}
Here we use a part of our case study to show the basics of our language. The model of the case study is presented in Fig.~\ref{fig:BBWModel}. Our case study is a Brake-By-Wire (BBW) with an Anti-lock Braking System (ABS). Here we only describe two classes \WheelClass\ and \WheelControllerClass\ which define the behavior of wheels and wheel controllers, respectively. The physical class \WheelClass\ has one known rebec \WheelClassController, which is the wheel controller of the wheel. It also has three variables \WheelClassTorque, \WheelClassSPD\ and \WheelClassTime\ which are respectively the amount of torque that must be applied during brake, the speed of the wheel and an auxiliary timer for periodic behavior of the wheel's sensor. In its \textit{initial} message server, the \WheelClassSPD\ is initialized to the given value and the active mode of the rebec is set to \WheelClassBrakeMode\ mode by using the \SetModeStatement\ statement. \WheelClass \ defines a setter message server \WheelClassSetTorque\ for updating the value of \WheelClassTorque. This class has one mode named  \WheelClassBrakeMode. This mode has a periodic behavior that every $50$ milliseconds, sends the value of  \WheelClassSPD \ to its wheel controller defined by \WheelClassController. For the periodic behavior the invariant $ \WheelClassTime <= 0.05$, the flow $\WheelClassTime' = 1 $ and the guard $ \WheelClassTime == 0.05 $ are defined. The flow equation for \WheelClassSPD\ variable is defined simply as $\WheelClassSPD' = -0.1 -\WheelClassTorque$. The constant $-0.1$ models the friction of the wheel with the road. In the actions of this mode, after reseting the timer value and sending the wheel's speed to its controller, the rebec's mode is again set to the \WheelClassBrakeMode\ mode if the wheel's speed is greater than zero.

%The only difference between these modes is the flow equation of \WheelClassSPD variable. In \WheelClassNoBrakeMode \ mode, we assume a small reduction rate in the value of \WheelClassSPD over time, but in \WheelClassBrakeMode\ mode the flow is defined simply as $\WheelClassSPD' = -\WheelClassTorque$. Note that in the actions of both behaviors, the rebec's mode is again set to the related mode.  

The ABS behavior is defined by the software class \WheelControllerClass. It has two known rebecs \WheelControllerClassWheel\ and \WheelControllerClassGlobalController, which are the controlled wheel of the controller and the global brake controller of the system, respectively. It also has three state variables \WheelControllerClassID, \WheelControllerClassSPD\ and \WheelControllerClassSliprate. The variable \WheelControllerClassID\ is the identifier of the wheel controller and is used to differentiate between multiple wheel controllers in the model, \WheelControllerClassSPD\ is the speed of the controlled wheel, and \WheelControllerClassSliprate\ is an auxiliary variable for calculating the slip rate. The variable \WheelControllerClassSliprate\ is used to determine whether the brake must be applied or not (this will be explained more in section \ref{sec::CaseStudy}). The \WheelControllerClassSPD\  variable gets updated by \WheelControllerClassSetWheelSPD\ message server. This message server also sends the wheel's speed along side the controller's identifier, to the \WheelControllerClassGlobalController. To calculate the brake torque for wheels, the global brake controller sends an \WheelControllerClassApplyTorque \ message to each \WheelControllerClass \ in the system. The corresponding message server has two parameters named \WheelControllerClassRequestedTorque\ and \WheelControllerClassVehicleSpeed\ which are the requested braking torque and the estimated vehicle speed, respectively. In this message server first the slip rate, denoted by the variable \WheelControllerClassSliprate, is calculated. Here the constant \WheelControllerClassWRAD\ denotes the radius of the wheel. After calculating the slip rate, if the value of \WheelControllerClassSliprate \ is greater than $0.2$, it sets its wheel's brake torque to zero by using the \WheelClassSetTorque\ message. Otherwise it sets the wheels brake torque to the requested torque.

In the main block, two rebecs of type \WheelClass\  and two rebecs of type \WheelControllerClass\ are instantiated. Here we explain one instantiation of each class. The term $\WheelClass~\WheelRRebec~(\AtWire~\WheelControllerRRebec):(1);$ is used to instantiate a rebec named $\WheelRRebec$ of type \WheelClass. The known rebec of $\WheelRRebec$ is assigned to $\WheelControllerRRebec$. The tag $\AtWire$ is used to indicate that the communications of this rebec to $\WheelControllerRRebec$ are via wire. The parameter $(1)$ is the parameter of the \textit{initial} message server of the rebec, which is the initial speed of the wheel. The term $\WheelControllerClass~\WheelControllerRRebec~(\AtWire~\WheelRRebec,~\AtCAN~\BrakeControllerRebec):(0);$ is used to instantiate a rebec named $\WheelControllerRRebec$ of type \WheelControllerClass. The known rebecs of $\WheelControllerRRebec$ are assigned to $\WheelRRebec$ and $\BrakeControllerRebec$. For the second known rebec, the tag $\AtCAN$ is used to indicate that the communications of this rebec to $\BrakeControllerRebec$ are via CAN. The parameter $(0)$ is the identifier of the wheel controller. In the priorities of CAN specification, the term $\WheelControllerRRebec~~\BrakeControllerRebec.\GlobalBrakeControllerSetSpeed~~3$, specifies that the priority of a message sent from $\WheelControllerRRebec$ to $\BrakeControllerRebec$ containing the message server name $\GlobalBrakeControllerSetSpeed$ is $3$. The term $\WheelControllerRRebec~~\BrakeControllerRebec.\GlobalBrakeControllerSetSpeed~~\textit{-\textgreater}~0.01$ in the delays of CAN specification, specifies that the delay of the same message is $0.01$.

%or \WheelControllerClassRequestedTorque \ is zero, \WheelControllerClass \ sets the mode of its wheel to \WheelClassNoBrakeMode, otherwise it sets \WheelControllerClassWheel's mode to \WheelClassBrakeMode \ while the wheel's \WheelClassTorque \ is set to \WheelControllerClassRequestedTorque. The \textit{initial} message server of this class is empty and is not shown for brevity.
%\WheelClassTorque \ to \WheelControllerClassRequestedTorque \  and then sets the mode of its controlling wheel to \WheelClassBrakeMode.

%\begin{figure}[hptb]
%	\centering
%	\lstinputlisting[language=HRebeca, multicols=2]{"../Examples/BrakeByWire-Part1-Simplified.rebec"}
%	\caption{The class definitions for \WheelClass\ and \WheelControllerClass}
%	\label{fig:exampleHA}
%\end{figure}

\def\NotSuspended{\neg v(suspended)}
\def\NotSuspendedSub[#1]{\neg v_#1(suspended)}

\def\gs{l}

\def\ds{{\it ss}}
\def\DS{{\it SS}}
\def\cs{{\it ps}}
\def\CS{{\it PS}}
\def\ns{{\it ns}}
\def\NS{{\it NS}}

\def\NSready{{\it r}}
\def\NSReady{{\it R}}
\def\NSbuffer{{\it b}}
\def\NSBuffer{{\it B}}
\def\es{{\it es}}
\def\ES{{\it ES}}
\def\gsDef{\gs=(\ds,\cs,\ns,\es)}
\def\ESpool{{\it p}}
\def\ESPool{{\it Pool}}
\def\ESevents{{\it evts}}
\def\ESevent{{\it evt}}
\def\ESEvents{{\it events}}
\subsection{Operational Semantics}\label{subsec::semantics}
Rebecs are executed concurrently in response to a physical mode being finished or by taking a message from their message queues. The actions of a physical mode are executed when its behavior is finished, and each message is processed by executing its corresponding message server. The execution of all the statements except the delay statement is instantaneous. 
%The statements of a message server are executed instantanously unless they are of delay statemet. 
%A delay statement suspends the executing rebec and generates an \emph{event} %responsible 
%by which the passage of time is modeled. Events are auxiliary concepts to make the semantics definition easier. An event, in the terminology of hybrid automata, is modeled by a real-valued timer whose timing behavior is defined by an appropriate flow and invariant.  To express the specific behavior upon expiration of an event, we employ different types of events. 
To model communication via CAN, a network entity is considered in the semantics which buffers the messages from the rebecs and 
delivers them one-by-one to the respective receivers based on the messages' priorities and delays specified in the model.
% Due to the delay in delivering a message, an event is generated by the network for the selected message, which upon its expiration, the message is inserted into the receiver's message queue. 
The message selection mechanism of CAN protocol is time consuming. We assume that this time is negligible. We abstract away from this time by considering the effect of the network entity when no actor can progress instantaneously. For communication via wire, the message is directly inserted into the receiver's message queue instantaneously. 
%The events generated due to the computation and message delays, are maintained by an explicit entity in the semantics. 

A Hybrid Rebeca model consists of the rebecs of the model and the network specification. A software rebec consists of the definitions of its variables, message servers and known rebecs. A physical rebec is defined like a software rebec plus the definitions of its modes. The network specification consists of the communication types of rebecs, which can either be CAN or wire, the message priorities and message delivery delays.

%A Hybird Rebeca model is defined as a tuple $(R_s,R_p,N)$ where $R_s$ and $R_p$ are the set of software and physical rebecs in the model, respectively, and $N$ is the network specification. In the remaining of the paper, we will use the set $R = R_s \cup R_p$ as the set of all rebecs in the model. A software rebec is defined with its unique identifier, variables, message servers and known rebecs. A physical rebec is defined like a software rebec plus the definition for its modes. The network specification consists of the communication types of rebecs, which can either be CAN or WIRE, the message priorities and message delays.

\begin{definition}[Hybrid Rebeca model]
A Hybrid Rebeca model is defined as a tuple $(R_s,R_p,N)$ where $R_s$ and $R_p$ are the set of software and physical rebecs in the model, respectively, and $N$ is the network specification. The set $R = R_s \cup R_p$ denotes the set of all rebecs in the model.

A software rebec $r_{s_{i}}\in R_s$ and physical rebec $r_{p_{i}}\in R_p$ with a unique identifier $i$, are defined by tuples $(i,V_{i},\MessageServers_{i},K_i)$ and $(i, V_{i},\MessageServers_{i},\Modes_{i},K_i)$, respectively, where $V_{i}$ is the set of its variables, $\MessageServers_{i}$ is the set of its message servers, $K_i$ is the set of its known rebecs, and %. A physical rebec $r_{p_{i}}$ with a unique identifier $i$, is defined as a tuple $(V_{i},\Me_{i},\Mo_{i},K_i)$ where $V_{i}$ is the set of its variable names, $\Me_{i}$ is the set of its message servers, 
$\Modes_{i}$ is the set of modes. %, and $K_i$ is the set of its known rebecs. 
%We define a message server as a tuple $(\MessageName,b)\in {\it Name} \times {\it Stmt}^*$ where $\MessageName$ is the name of the message server and $b$ is the body of the message server which is a sequence of statements. For simplicity we ignore the message server parameters here. A mode $\ModeIden$ is defined as a tuple $(\ModeName,i,f,g,a) \in  {\it Name} \times {\it Expr} \times Expr \times Expr \times \Statement^*$ where $\ModeName$ is the name of the mode, $i$, $f$, $g$  and $a$ are respectively invariant, flows, guard and actions of the mode. We define a message as a tuple $(sender, \MessageName, receiver) \in R\times {\it Name} \times R$ where $sender$ is the sending rebec, $\MessageName$ is the name of the message server in the receiver and $receiver$ is the receiving rebec. Let ${\it Msg}$ denote the set of all messages, ranged over by $M$.

A network specification is defined as a tuple $N = ({\it conn}, {\it netPriority},{\it netDelay})$ where ${\it conn}$ is a partial function $R\times R \rightarrow \{{\it Wire},{\it CAN}\}$  which defines the one way connection type from a rebec to another rebec, %. $can$ is the CAN specification which defines the 
${\it netPriority}: {\it Msg} \rightarrow \mathbb{N}$ and  ${\it netDelay}: {\it Msg} \rightarrow \mathbb{R}$ define the priority and the network delivery delay for a message, respectively. ${\it Msg}$ denotes the set of all messages in the model.
\end{definition}

\begin{definition}[Message]
	A message  is defined as a tuple $(sender, \MessageName, receiver) \in R\times {\it Name} \times R$ where $sender$ is the sending rebec, $\MessageName$ is the name of the message server in the receiver, and $receiver$ is the receiving rebec.
\end{definition}

%\subsection{Intermediate Labeled Transition System}
\def\EventResume{\textit{Resume}}
\def\EventEnqueue{\textit{Transfer}}
\def\EventEffectFunction{\textit{effect}}
\def\AuxGuard{\textit{guard}}
\def\AuxInvariant{\textit{invariant}}
\def\AuxActions{\textit{actions}}
\def\AuxFlows{\textit{flows}}

The operational semantics of a Hybrid Rebeca model is defined as a monolithic hybrid automaton. The semantics could be defined in a compositional way by providing a translation for constitutive elements of a hybrid rebeca model. By composing the translations with the operational semantics of hybrid automata, the final model can be derived. However, this approach will lead to many real-valued variables which reduces analyzability of the resulting model.

\begin{definition} [Hybrid automaton for Hybrid Rebeca model]
Given a Hybrid Rebeca model  $\mathfrak{H}=(R_s,R_p,N)$, its formal semantics based on hybrid automata is defined by $H_{\mathfrak{H}}=(\HALoc,\HAVar,\HALab, \Rightarrow, \HAFlws, \HAInv, \HAInit)$, where $\HAVar$ is the set of all continuous variables in the model (variables of types \textit{float} or \textit{real}), $\HALab$ is the set of labels which is empty as we generate a monolithic hybrid automaton. %, the set of labels are empty.  
The set of locations $\HALoc$, transitions $\Rightarrow$, flows $\HAFlws$, invariants $\HAInv$, and initial conditions $\HAInit$ are defined in the following.
\end{definition}
 % semantics of Hybrid Rebeca as follows

\subsubsection{Locations}
Each location has four entities: the states of software rebecs, physical rebecs, network, and pending event list. The state of a software rebec consists of the valuation of its discrete variables, the state of its message queue and a program counter. The program counter points to a statement that the rebec must execute. The state of a physical rebec consists of its active mode, the state of its message queue and a program counter. The state of a physical rebec does not contain any valuation since discrete variables are not defined for physical rebecs and the continuous variables are handled in the hybrid automaton. A software rebec has the notion of being suspended (due to the execution of a delay statement). The suspension status is maintained by a reserved variable in the valuation of the rebec. Delay statements are not allowed in physical rebecs.

\begin{definition}[State of a rebec]
	The state of a software rebec is denoted by the tuple $(v,q, c)$ where $v$ is the valuation of its variables, $q$ is the message queue of the rebec,  %which is a sequence of messages 
	and $c$ denotes the program counter. The state of  a physical rebec is a tuple of the form $(\ModeIden, q, c)$ where $\ModeIden$ is the active mode and $q$ and $c$ are defined as in the software rebec's state.
\end{definition}

The network state, which is the state of the CAN network, consists of the buffered messages in the network and the status of the network which indicates that the network is busy sending a message or is ready to send one. 
\begin{definition}[State of network]
	The network state is defined by the pair $(\NSBuffer,\NSready)$, where $\NSBuffer$ is the network buffer and %which is a sequence of messages and 
	the boolean flag $\NSready$ indicates the status of the network, which can be ready or busy.
\end{definition}

 %Each delay statement is modeled 

The forth entity, pending event list, represents the sequence of pending events. Events are used for time consuming actions. For a time consuming action an event is stored, to be triggered at the time that the action is over.
Two types of events are defined in the semantics of Hybrid Rebca: \EventResume\ and \EventEnqueue. A pending event with event \EventResume\ is generated and inserted into the pending event list when a delay statement is seen in Rebeca model, and the corresponding rebec is suspended. To model the passage of time for the delay statement, a timer variable is used by the pending event. After the specified delay has passed, the event is triggered, and consequently the behavior of the given rebec is resumed by updating the suspension status of the rebec.  %Upon executing a send statement (to a rebec connected via CAN), the message is inserted into the network buffer. 
A pending event with event \EventEnqueue\ is generated when a message from the network buffer is chosen to be sent to its receiver. A timer is assigned to model the message delivery delay, and the pending event is inserted into the pending event list. Upon triggering of a \EventEnqueue\ event, the specified message is  enqueued in the receiver's message queue, and the network status is set to ready which means the network is ready to send another message.

\begin{definition}[Pending event]
	A pending event is a tuple $(d,e,t)$ where $d$ is the delay of the event $e$ and $t$ is a timer variable that is assigned to this event. The event $e$ can either be a \EventResume\ or \EventEnqueue\ event. The timer variable is used for defining the timing behavior for the delay of the pending event. The event is triggered (and executed) after $d$ units of time after the pending event is created.
\end{definition}

\def\RuleNameMessageTakeFIFO{Message Take (FIFO)}
\def\RuleNameContAssign{Continuous  Assignment}
\def\RuleNameDiscAssign{Discrete  Assignment}
\def\RuleNameCondFalse{Conditional False}
\def\RuleNameCondTrue{Conditional True}
\def\RuleNameCondUnknown{Conditional Unknown}
\def\RuleNameDirectMessage{Direct Message Send}
\def\RuleNameCANMessage{CAN Message Send}
\def\RuleNameChangeMode{Set Mode}
\def\RuleNameDelay{Delay Statement}
\def\RuleNameCANSchedule{CAN Schedule}
\def\RuleNameEventExpiration{Event Expiration}
\def\RuleNameModeExpiration{Mode Expiration}

\subsubsection{Transitions}
We define two general types of transitions: urgent and nonurgent transitions. The urgent transitions are further divided to message, statement and network transitions which are respectively shown as \MessageTransition, \StatementTransition\ and \NetworkTransition. The nonurgent transitions are shown as \NonurgentTransition. We use these transitions to differentiate between different types of actions.
Message transitions are only for taking a message. Statement transitions are for executing the statements. A network transition chooses a message from the buffer of the network to be sent. This transition is only about choosing the message and not the act of sending. Non-urgent transitions are used to model the passage of time. %continuous behaviors of the system. 
These transitions include the behaviors of physical actors' active modes and pending time of events since they are time consuming. % Physical transitions are taken when the physical behavior of its location is finished. 

An ordering is defined among these transitions. The ordering is \MessageTransition \ $=$ \StatementTransition\ $>$  \NetworkTransition \ $>$ \NonurgentTransition. Whenever a higher order transition is enabled in a location, no lower order transition can be taken in that location. The semantics of the actions in Hybrid Rebeca are defined using these transitions on the locations. In the following we define these transitions.

\paragraph{Message Transitions:} Message transitions define the act of taking a message.  A message transition can take place whenever a rebec is not suspended. A rebec is suspended when it executes a delay statement. Let the tuples $(v,q, c)$ and  $(\ModeIden, q, c)$ denote the state of a software rebec and a physical rebec respectively. The message transition is defined as follows:
\begin{itemize}
	\item \textit{Taking a Message:} A rebec takes a message from the head of its message queue $q$, whenever the rebec has no statement to execute. When a message is taken, the program counter $c$ is updated to point to the beginning of the corresponding message server, and the message is removed from the message queue $q$.
\end{itemize}

\paragraph{Statement Transitions:} Statement transitions define the act of executing the statements. Like message transitions, a statement transition can take place whenever a rebec is not suspended. Consider the tuples $(v,q, c)$ and  $(\ModeIden, q, c)$ as the states of a software rebec and a physical rebec respectively. The statement transitions include the followings:
\begin{itemize}
	\item \textit{Assignment Statement:} This statement has two cases. When assigning to a discrete variable, the value of the variable is update in the valuation $v$ of the rebec. When assigning to a continuous variable, since its value is not determined (it may depend on the  continuous behaviors), the assignment is transfered over to the transition to be handled by the resulting hybrid automaton.
	\item \textit{Conditional Statement:} This statement has two cases. If the value of the condition is determined, the program counter $c$ is updated to point to the appropriate statement block. If the value of the condition is not determined (because of continuous variables used in the condition), both possible paths are considered by creating two separate transitions. The condition and its negation act as the guards for these transitions.
	\item \textit{Send Statement:} This statement, depending on the communication type, has two behaviors. When the communication is via wire, the message is directly added to the receiver's message queue. When the communication is via CAN, the message is added to the CAN buffer to be handled by the CAN‌ behavior.
	\item \textit{Delay Statement:} This statement suspends the software rebec by updating  the corresponding variable (suspension status) in the valuation $v$ and creates a pending event $(d,\EventResume,t)$ for resuming the rebec after $d$ units of time. $d$ is the delay specified in the delay statement and $t$ is a timer variable.
	\item \textit{Set Mode Statement:} This statement changes the active mode $\ModeIden$  of the physical rebec to the specified mode.
\end{itemize}

\paragraph{Network Transitions:} 
These transitions define the behavior of the CAN network which only includes the behavior of choosing a message from the network buffer to be sent.  Since network transitions have a lower priority than message and statement transitions, this makes the choosing behavior to happen only when no rebec can progress instantaneously. Let $(\NSBuffer,\NSready)$ be the network state. The choosing transition is as follows: 
\begin{itemize}
	\item \textit{Choosing a Message:} For this behavior, the message with the highest priority is removed from the network buffer $\NSBuffer$, a pending event $(d,\EventEnqueue,t)$ for sending the message is created, and the flag $\NSready$ of the network is updated to indicate that the network is busy. The delay $d$ for the created pending event is the network delay of the message.

\end{itemize}

\paragraph{Nonurgent Transitions:} 
Nonurgent transitions are used to define the end of active physical modes and triggering pending events. These transitions are defined only when no urgent transition is possible. These transitions are as follows: 
\begin{itemize}
	\item \textit{End of an Active Mode:} For a physical rebec $(\ModeIden, q, c)$ if $\ModeIden$ is not \None, the guard of the active mode $\ModeIden$ is transfered to the transition, and the program counter $c$ of the rebec is updated to point to the actions of this active mode, and the active mode is set to \None.
	\item \textit{Triggering of an Event:} For a pending event  $(d,e,t)$, the guard $t==d$ is defined on the transition where $t$ and $d$ are the timer and the delay of the pending event respectively. The event $e$ is executed as a result of this transition and the pending event is removed from the pending event list.  
\end{itemize}

\def\UrgFlow{\textit{urg}'=1}
\def\UrgInv{\textit{urg} \le 0}
\def\ModeFlows{\textit{ModeFlows}}
\def\EventFlows{\textit{EventFlows}}
\def\ConstantFlows{\textit{FloatFlows}}

\def\ModeInvs{\textit{ModeInvs}}
\def\EventInvs{\textit{EventInvs}}

\subsubsection{Flows and Invariants}
%To define the semantics of Hybrid Rebeca we need to define the continuous behaviors and instantaneous behaviors of the model in hybrid automata. There are two kinds of continuous behaviors in the model, behaviors regarding physical rebec modes and behaviors regarding events. Physical modes have all the necessary information in themselves and events are simply a timing behavior. 
To define flows and invariants for each location we need to consider continuous and instantaneous behaviors separately. There are two kinds of continuous behaviors in the model, behaviors regarding physical rebecs' modes and behaviors regarding pending events. Physical modes have all the necessary information in themselves and the pending events have simple timing behaviors. The functions $\AuxFlows(r_p,\ModeIden)$ and  $\AuxInvariant(r_p,\ModeIden)$ return the flow and invariant of a mode $\ModeIden$ in a physical rebec $r_p$ respectively. Instantaneous behaviors should be executed without allowing the time passage. So time should not be passed when the system resides in the source locations of such transitions, called \emph{urgent} locations.
\begin{definition}[Urgent location flow and invariant]
	 A possible implementation for an urgent location  is ${\it urg}'=1$ as flow and ${\it urg}<=0$ as invariant, where  ${\it urg}$ is a specific variable. Note that in this method, this new variable must be added to the set \HAVar\ of the hybrid automaton. Also the assignment ${\it urg} = 0$ must be added to all incoming transitions to an urgent location. The defined invariant prevents the model from staying in the location as the value of $\it urg$ will be increased by the defined flow. 
\end{definition}

If a location is urgent, the urgency flow, as defined above, should be set as its flows.  In case a location is not urgent, it inherits the flows of all physical rebecs' active modes, the flows for timers of pending events, and a flow of zero for each float variable to freeze its value. The flow of a pending event is simply defined as $t'=1$ where $t$ is the timer variable of the pending event. %The \HAFlws \ function of the hybrid automaton is defined as:
%
%\[\HAFlws(\gs)=\begin{cases}
%\UrgFlow,& \text{if $\gs$  is urgent} \\
%\ModeFlows(\gs) \bigcup \EventFlows(\gs) \bigcup \ConstantFlows(\gs) ,& \text{otherwise}
%\end{cases}\]
%
%\[ \begin{array}{ll}
%\ModeFlows(\gs)  &= \sideset{}{}\bigcup_{x \in \text{physical rebecs}} \AuxFlows(x,\ModeIden)\ \text{where $\ModeIden$ is the active mode of $x$} \\
%\EventFlows(\gs) &= \sideset{}{ }\bigcup_{(d, e, t) \in \text{pending events of $\gs$}} t'=1 \\
%\ConstantFlows(\gs) &= \sideset{}{}\bigcup_{v \in \text{float variables}} v'=0\ 
%\end{array}\] 

Similarly, if a location is urgent, its invariant is set to urgency invariant, otherwise it inherits the invariants of all physical rebecs' active modes, and the invariants for corresponding pending events' timers. The invariant of a pending event is defined as $t \le d$ where $t$ and $d$ are the timer variable and the delay of the pending event, respectively. %The \HAInv\ function of the hybrid automaton is defined as:

%\[\HAInv(\gs)=\begin{cases}
%\UrgInv,& \text{if $\gs$  is urgent} \\
%\ModeInvs(l) \bigcup \EventInvs(l)  ,& \text{otherwise}
%\end{cases}\]
%
%
%\[ \begin{array}{ll}
%\ModeInvs(\gs)  &= \sideset{}{}\bigcup_{x \in \text{physical rebecs}} \AuxInvariant(x,\ModeIden)\ \text{where $\ModeIden$ is the active mode of $x$} \\
%\EventInvs(\gs) &= \sideset{}{ }\bigcup_{(d, e, t) \in \text{pending events of $\gs$}}t \le d \\
%\end{array}\] 

\subsubsection{Initial Location and Initial Condition}
For the initial location $\gs_0$, we initialize all discrete variables of rebecs to the value zero. Furthermore, the initial message for each instantiated rebec, is put into its message queue. We also set the value of all continuous variable to zero in the initial condition of the initial location. %The function \HAInit\ of the hybrid automaton is defined as below.

%\[\HAInit(\gs)=\begin{cases}
%\sideset{}{}\bigcup_{v \in \text{continuous variables}} v=0\, & \text{if $\gs = \gs_0$} \\
%\emptyset  ,& \text{otherwise}
%\end{cases}\]

\subsection{Technical Details}
For simplicity some details were omitted from our semantics. Here we describe these details informally.

\textbf{Limited Size for Message Queues:} In the semantics of Hybrid Rebeca, the message queues of rebecs are considered unbounded. But in practice a specific size must be specified for message queues of rebecs.

\textbf{Message Arguments:} To incorporate message arguments, we must consider discrete and continuous arguments separately. For discrete arguments, since their values are known in the state, that value is included in the message and when the message is taken, its arguments are added to the rebec's valuation. When the execution of the message is finished, the arguments are removed from the valuation. For continuous arguments, the values are not generally determined during the translation to hybrid automata, so it's not possible to send the value within the message. 
%Since such values may change over time, it is not possible to send the references to the original variables neither (similar to the concept of call-by-reference in programming languages). 
To this aim, a non-evolving auxiliary variable is used. Before sending a message, each continuous argument is assigned to an auxiliary variable (by using continuous variable assignment). Then, a reference to the auxiliary variable is included in the message. When the message is taken, for each continuous argument, an assignment from the auxiliary variable to its respective parameter variable is implicitly executed. % all statements %on the corresponding argument 
%will be defined on the auxiliary variables.

\textbf{Continuous Variable Pools:} When creating a new event (for a delay statement or for sending a message from CAN), a new timer is assigned to each event. But in hybrid automata all continuous variables must be defined statically. To handle this, variable pools with fixed sizes are used. There are two variable pools in our semantics, one for timer variables and one for the auxiliary variables of message arguments (as mentioned above). The size of the variable pools affects the behavior of the model. A small size will lead to an incomplete model, and a large size will lead to a huge model which can not be easily analyzed. % An upper bound can be found for the size of these pools by static inspection of the model. For the timer pool, it is the number of rebecs plus one (for the network entity). For the float pool, it is sum of the size of rebec's queues times the maximum number of arguments.

\section{Case Study}\label{sec::CaseStudy}
We demonstrate the applicability of our language on a simplified Brake-by-Wire (BBW) system with Anti-lock Braking System (ABS) \cite{KangEMSSP13,MarinescuMS16,FilipovikjMMSLL16}. In a BBW system instead of using mechanical parts, braking is managed by electronic sensors and actuators. In ABS, the safety is increased by releasing the brake based on the slip rate to prevent uncontrolled skidding.
  
\begin{figure}[H]
	\centering
	\includegraphics[width=8.3cm]{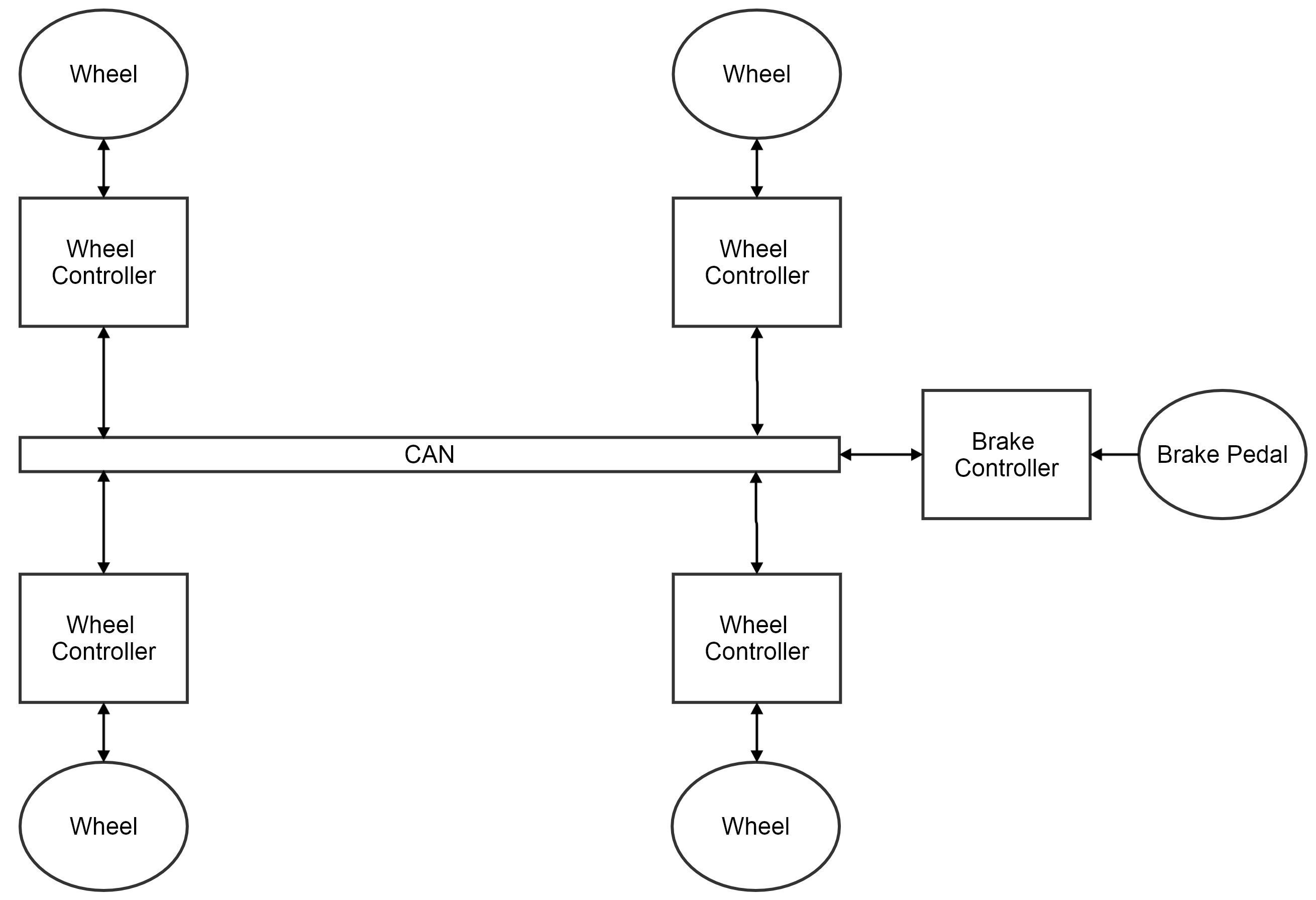}
	\caption{The schematic of the BBW system. The physical components are shown as ellipses and computational components are shown as rectangles. }
	\label{fig:BBWSystem}
\end{figure}

In this system, the brake pedal sensor calculates the brake percentage based on the position of the brake pedal. A global brake controller computes the brake torque and sends this value to each  wheel controllers in the vehicle. Each wheel controller monitors the slip rate of its controlled wheel and releases the brake if the slip rate is greater than $0.2$. There is a nonlinear relationship between the friction coefficient of the wheel and the slip rate. When the slip rate is between zero and around $0.2$, any increase in the slip rate increases the friction coefficient, but after $0.2$, further increase in the slip rate, leads to a reduction in the friction coefficient. For this reason when the slip rate is greater than $0.2$, no brake will be applied to the wheel. In this system, each pair of wheel and its wheel controller are connected directly by wire. Furthermore, the brake pedal sensor sends the brake percentage value to the global brake controller through wire. All other communications are done through a shared CAN network. A representation of the system is shown in Figure \ref{fig:BBWSystem}.

\subsection{Model Definition}
The model is defined in Fig.~\ref{fig:BBWModel}. Note that For simplicity, we have considered two wheels in our model. The model consists of 5 classes. \WheelClass \ and \WheelControllerClass \ classes were defined previously in \ref{subsec::example}. \WheelClass\ class models the sensors and actuators of the wheel. It periodically sends the speed of the wheel to the controller and defines the effect of braking on the wheel speed. \WheelControllerClass\ class defines the behavior of the wheel controller. It monitors the slip rate of the wheel and decides to apply the brake based on its value.

\BrakePedalClass \ class defines the behavior for the brake pedal. Here we assume a simple behavior where the brake percentage is increased by a constant rate until it reaches a predefined max percentage. The class have one known rebec \BrakePedalGlobalController\ which is the global brake controller. It defines four state variables \BrakePedalBrakePercent, \BrakePedalMaxBrakePercent, \BrakePedalTimer\ and \BrakePedalRate\ which are the brake percentage, maximum brake percentage, an auxiliary timer variable and a variable that defines the rate for the brake percentage. In the \textit{initial} message server, the values of the initial and maximum brake percentage are initialized with the given values and the rate variable is set to $1$ and the active mode of the rebec is set to \BrakePedalBraking. \BrakePedalBraking\ mode defines a periodic behavior where the value of \BrakePedalBrakePercent \ is sent to \BrakePedalGlobalController\ and the brake percentage is increased by the rate defined by \BrakePedalRate. In the actions of this behavior, if the brake percentage is equal or greater than \BrakePedalMaxBrakePercent, the rate variable \BrakePedalRate\ is set to zero to stop the brake percentage from changing by time.
%the rebec stays in the current mode, otherwise it goes to \BrakePedalConstantBrake\ mode. \BrakePedalConstantBrake\ mode's behavior is like \BrakePedalIncreasingBrake\, except that a zero flow is defined for the \BrakePedalBrakePercent.
 
\GlobalBrakeControllerClass \ class is the global brake controller and has the responsibility of delegating the brake torque to wheel controllers. It defines two known rebecs for each wheel controller named \GlobalBrakeControllerWCTLRR\ and \GlobalBrakeControllerWCTLRL. \GlobalBrakeControllerClass \ class has five state variables for the speed of the right and left wheels, the brake percentage from the brake pedal, the global torque calculated from the brake percentage and the estimated vehicle speed. In the message server \GlobalBrakeControl, first the estimated speed of the vehicle is computed based on the speed of individual wheels and the desired brake torque is calculated based on the brake percentage. Here we simply assume that the brake percentage is equal to the brake torque. Then, the estimated speed and global torque are sent to each wheel controller. The \textit{initial}, \GlobalBrakeControllerSetSpeed\ and \GlobalBrakeControllerSetBrakePercent\ message servers are omitted for brevity. The \GlobalBrakeControllerSetSpeed\ message server updates the correct wheel speed variable based on the input identifier. The message server \GlobalBrakeControl\ must be executed periodically, so an auxiliary \ClockClass \ class is used to periodically send a  \GlobalBrakeControl \ message to \GlobalBrakeControllerClass. 

In the main block, all necessary rebecs are instantiated. The wheels \WheelRRebec\ and \WheelLRebec\ are wired to their respecting wheel controllers by using the tag @\WireCon. Both wheels are initialized with the speed of $1$\footnote{As the properties to be verified do not depend on the value of the speed, to minimize the analysis time, this value has been chosen.}. The wheel controllers \WheelControllerRRebec\ and \WheelControllerLRebec\ are connected to their corresponding wheels by wire and are connected to the global brake controller through CAN by using the tag @\CANCon. Identifiers of $0$ and $1$ are given to rebecs \WheelControllerRRebec\ and \WheelControllerLRebec\ as initial parameters, respectively. The brake controller \BrakeControllerRebec\ is connected to both wheel controllers through the CAN network and the brake \BrakeRebec\ is initialized with the brake percent $60$ and maximum brake percent of $65$. Both brake \BrakeRebec\ and clock \ClockRebec\ are connected to \BrakeControllerRebec\ by wire. There are four CAN messages in the model. The brake controller \BrakeControllerRebec\ sends \WheelControllerClassApplyTorque\ message to the wheel controllers \WheelControllerRRebec\ and \WheelControllerLRebec. The wheels \WheelRRebec\ and \WheelLRebec\ send \GlobalBrakeControllerSetSpeed\ message to \BrakeControllerRebec\ respectively. A higher priority is defined for \WheelControllerClassApplyTorque\ messages. Note that a lower number indicates a higher priority. The network delay of all four CAN messages is specified as $10$ milliseconds.

\def\HAMonitorTime{\textit{monitor\_time}}
\def\HAMonitorRTime{\textit{mntrR\_time}}
\def\HAMonitorLTime{\textit{mntrL\_time}}
\def\HAWheelTorque{\textit{wheel\_torque}}
\def\HAWheelRTorque{\textit{wheelR\_torque}}
\def\HAWheelLTorque{\textit{wheelL\_torque}}
\def\HAWheelControllerSliprate{\textit{wctlr\_slprate}}
\def\HAWheelControllerRSliprate{\textit{wctlrR\_slprate}}
\def\HAWheelControllerLSliprate{\textit{wctlrL\_slprate}}
\subsection{Analysis and Verification}
For the analysis of this model the queue size of \BrakeControllerRebec\ is set to $4$, the queue sizes of both \WheelControllerRRebec\ and \WheelControllerLRebec\ is set to $2$, and for other rebecs the queue size is set to $1$. The size of timer variable pool is set to $1$ and the size of arguments variable pool is set to $11$. 
The hybrid automaton derived from the model consists of $10097$ locations and $25476$ transitions. We use SpaceEx \cite{frehse2011spaceex} tool to verify our model.  Note that the slip rate equation used in the model is not supported by SpaceEx, since it's a nonlinear equation. We simplified this equation for analysis. By specifying a set of forbidden states, safety properties can be verified by SpaceEx. We developed an initial tool\footnote{The implementation can be found in https://github.com/jahandideh-iman/HybridRebeca} for our language that automatically translates a Hybrid Rebeca model to a SpaceEx model based on the semantics of our language. We verified three properties for our case study. The first property is ``design-fault freedom". Here by design-fault we mean a fault caused by following situations:‌ exceeding the capacity of a message queue, running out of pooled variables, and having messages with same priority in CAN buffer. We assume the message with the highest priority must always be unique in CAN buffer. Note that in practice, this property must be implicitly  checked for all models in Hybrid Rebeca, but for now it must be manually verified by SpaceEx. The second property is a timing constraint. This property states that the time between the transmission of the brake percentage from the brake pedal, and its reaction by wheel actuators, must not exceed $0.2$ seconds. The third property states that whenever the slip rate of a wheel exceeds $0.2$, the brake actuator of that wheel must immediately be released.

\begin{figure}[H]
	\centering
\begin{lstlisting}[language=HRebeca, multicols=3]
physicalclass Wheel{
	knownrebecs {WCtlr ctlr;}
	statevars {float trq; real spd; real t;}
	msgsrv initial(float spd_){
		spd = spd_;
		setmode(Rolling);
	}
	msgsrv setTrq(float trq_){
		trq = trq_;	
	}
	mode Rolling{
		inv(t <= 0.05){
			t' = 1;
			spd' = -0.1-trq;
		}
		guard(t == 0.05){
			t = 0;
			ctlr.setWspd(spd);
			if(spd > 0)
				setmode(Rolling);
}}}
	
softwareclass WCtlr{
	knownrebecs {Wheel w; BrakeCtlr bctlr;}
	statevars {int id; float wspd; float slprt;}
	msgsrv initial(int id_){
		id = id_;
	}
	msgsrv setWspd(float wspd_){
		wspd = wspd_;
		bctlr.setWspd(id,wspd);
	}
	msgsrv applyTrq(float reqTrq, float vspd){
		if(vspd == 0)
			slprt = 0;
		else
			slprt = (vspd - wspd * WRAD)/vspd;
		if(slprt > 0.2)
			wheel.setTrq(0);
		else
			wheel.setTrq(reqTrq);
}}

physicalclass Brake{
	knownrebecs {BrakeCtlr bctlr;}
	statevars {real bprcnt; real t; float mxprcnt; float r}
	msgsrv initial(float bprcnt_, float mxprcnt_){
		bprcnt = bprcnt_;
		mxprcnt = mxprcnt_;
		r = 1;
		setmode(Braking);
	}
	mode Braking{
		inv(t <= 0.05){
			t' = 1;
			bprcnt' = r;
		}
		guard(t == 0.05){
			t = 0;
			bctrl.setBprcnt(bprcnt);
			if(bprcnt>=mxprcnt)
				r = 0;
			setmode(Braking);
}}}

softwareclass BrakeCtlr{
	knownrebecs{
		WCtlr wctlrR;WCtlr wctlrL;}
	statevars {float wspdR;float wspdL;float bprcnt;float gtrq;float espd;}
	msgsrv control(){
		espd = (wspdR + wspdL)/2;
		gtrq = bprcnt;
		wctlrR.applyTrq(gtrq, espd);
		wctlrL.applyTrq(gtrq, espd);
	}
	// Setters for wspdR, wspdL and bprcnt
	...
}

physicalclass Clock{
	knownrebecs {BrakeCtlr bctlr;}
	statevars {real t;}
	msgsrv initial(){
		setmode(Running)
	}
	mode Running(){
		inv(t <= 0.05){
			t' = 1;
		}
		guard(t == 0.05){
			t = 0;
			bctlr.control();
			setmode(Running);
}}}

main {
	Wheel wR (@Wire wctlrR):(1);
	Wheel wL (@Wire wctlrL):(1);
	WCtlr wctlrR (@Wire wR, @CAN bctlr):(0);
	WCtlr wctlrL (@Wire wL, @CAN bctlr):(1);
	BrakeCtlr bctlr (@CAN wctlrR, @CAN wctlrL):();
	Brake brake(@Wire bctlr):(60,65);
	Clock clock(@Wire bctlr):();
	
	CAN{
		priorities{
			bctlr	wctlrR.applyTrq		1;
			bctlr	wctlrL.applyTrq		2;
			wctlrR	bctlr.setWspd		3;
			wctlrL	bctlr.setWspd		4;
		}
		delays{
			bctlr	wctlrR.applyTrq		0.01;
			bctlr	wctlrL.applyTrq		0.01;
			wctlrR	bctlr.setWspd		0.01;
			wctlrL	bctlr.setWspd		0.01;
}}}
\end{lstlisting}
	\caption{The model definition of the BBW model.}
	\label{fig:BBWModel}
\end{figure}

For the first property, a specific location, called $\textit{Fault}$, is created for the mentioned situations, and during the semantic derivation, occurrences of such design-faults are handled by generating a transition to the specified location $\textit{Fault}$. The verified forbidden condition in SpaceEx for this property is $\textit{loc}() == \textit{Fault}$, where %$\textit{Deadlock}$ is the name of the deadlock location. In SpaceEx
 the term $\textit{loc}()$ specifies the current location in SpaceEx. 

The second property can not directly be specified with a set of forbidden states, since there is no direct concept of time in hybrid automata. To this aim, a monitor class is added to the model to measure the time between two events. Here the events are sending the brake percentage from the brake pedal to the global brake controller and processing the received brake torque in the wheels. The monitor class is a simple physical class with one physical mode and two stop and start message servers. The physical mode tracks the time and the message servers are used to stop and start the tracking. Note that in the start message server the tracked time is reset. 

For the second property, one monitor rebec is instantiated and is wired to the brake pedal and one of the wheels. 
%and each one is wired to one wheel and the brake pedal. %(Note that since there are two wheels, two monitor rebecs are needed. To verify both wheels, the monitor rebec given to the brake pedal must be changed). 
A start message is sent in the actions of the \BrakePedalBraking\ mode of the brake pedal after the brake percentage is sent to the brake controller and a stop message is sent by \WheelClassSetTorque\ message server of the wired wheel. In SpaceEx the forbidden condition $\HAMonitorTime > 0.2 $ is verified where $\HAMonitorTime$ is the name of the monitor's timer in the resulted hybrid automaton.

The third property may seem to be straightforward, but because the semantics of our language is fine-grained, the request of releasing the brake and the actual act, take place in different locations, even though the time does not advance between these locations because they are urgent locations. For this property, the monitor class is used again. This time, a monitor rebec is wired to one of the wheels and its wheel controller. A start message is sent from the wheel controller when the slip rate is greater than $0.2$ and a stop message is sent like the second property. By using the monitor rebec, the states which the brake is not released and time has not progressed while the slip rate is greater than $0.2$ can be considered safe. The verified forbidden condition in SpaceEx is\[\begin{array}{l}
\HAWheelControllerSliprate > 0.2 \,\wedge\, \HAWheelTorque > 0 \,\wedge\, \HAMonitorTime > 0
\end{array}
\] where $\HAWheelControllerSliprate$, $\HAWheelTorque$ and $\HAMonitorTime$ are the names of the slip rate of the wheel, the brake torque of the wheel and the timer of the monitor, respectively, in the resulted hybrid automaton.

The resulted hybrid automaton for the first property has $10097$ locations and $25476$ transitions, which is huge for verification purposes. This huge size stems from the fine-grained semantics of our language. But most of these locations are urgent locations where time does not advance and can be aggregated for the properties mentioned here. After aggregating these urgent locations, the size of the resulting hybrid automaton is reduced to $21$ locations and $1148$ transitions. The aggregation process is implemented in our tool. The three properties are verified on their respective reduced hybrid automaton. The verification result of these properties are provided in the Table \ref{tab:verificationResult}.

\def\Property{\textbf{Property}}
\def\DerivedHA{\textbf{Derived HA}}
\def\GenTime{\textbf{Gen Time}}
\def\AbsHA{\textbf{Red HA}}
\def\VerifResult{\textbf{Verif Result}}
\def\VerifDuration{\textbf{Verif Duration}}
\begin{table}
	\centering
\begin{tabular}{|c|c|c|c|c|c|c|c|}
	\hline 
	\Property 	& \multicolumn{2}{c|}{\DerivedHA }	& \GenTime\ (s) &  \multicolumn{2}{c|}{ \AbsHA } & \VerifResult & \VerifDuration\  (s) \\ 
	\hline 
	Design-fault Freedom	& 10097  	& 25476  & 12 & 21 & 1148  & Passed & 3705 \\ 
	\hline 
	Reaction Time		& 16317 	& 42976 & 20 & 21 & 1168 & Passed & 7521 \\ 
	\hline 
	Brake Release		& 54097 	& 175036 & 64 & 21 & 1168  & Passed &  3541 \\ 
	\hline 
\end{tabular}
  \caption{The verification result of the case study. Legends: \Property: verified property, \DerivedHA: derived hybrid automaton size where the first and second columns are the number of locations and transitions, respectively, \GenTime: duration of hybrid automaton generation in seconds, \AbsHA:‌ reduced hybrid automaton size, \VerifResult: result of verified property, \VerifDuration:‌ duration of verification in seconds. }\label{tab:verificationResult}
\end{table}

\section{Related Work}\label{sec::related}
There are some frameworks for modeling and analyzing cyber-physical systems. Some of these frameworks rely on simulation for analysis and others offer formal verification.  

Ptolemy II \cite{ptolemaeus2014system} relies on simulation for analysis and is a framework that uses the concept of \textit{model of computation} (MoC) which defines the rules for concurrent execution of components and their communications. Ptolemy supports many models of computation like process networks, discrete events, dataflow and continuous time. Heterogeneous model can be made by nesting these models of computations in a hierarchical structure. As far as we know there is no formal semantics for the hybrid models of Ptolemy framework to enable formal verification.
%TODO::actor model can be specified as an extension of discrete event 
In \cite{CicirelliNS18} an agent-based and control centric methodology is presented for development of CPSs. This approach includes all development stages of a system from analysis by simulation to the execution of the final system. For the modeling phase concepts like actors, message, actions, processing units and environmental gateway are presented in this methodology. The message passing among actors is asynchronous and the computations of the model take place in the actions that are submitted to the processing units by the actor for execution. The environmental gateway is used for abstracting the physical processes where in later stages is replaced by the real entities. This approach relies on simulation to analyze a system, and no formal analysis is supported. In \cite{TowardsModularCPS} a modular approach for specifying and validating CPSs using rewriting logic-based technique is purposed. In this work a CPS is described as a linear hybrid automata in rewriting logic where the components of the system communicate asynchronously. Timed hybrid Petri nets \cite{DavidA01} can also be used to model hybrid systems and CPSs. For analysis of these hybrid Petri nets in \cite{DavidA01} a translation to hybrid automata is presented. However, Petri-net based approaches prohibit modular specification of systems. The framework of \cite{Merro} provides a hybrid process calculus tailored for modeling CPSs and analyzing their security properties \cite{attacks,metric}. In this approach network governing the interactions between physical and cyber entities is not addressed.

\section{Discussions}\label{Sec:Discussions}
In Section \ref{sec::CPSActorModel} we presented our extended actor model for cyber-physical systems. In our model, the software and physical actors are separated and modes are added to physical actors for specifying the continuous behaviors. The separation of software and physical rebecs prevents the interference of continuous behaviors with software behaviors. In Rebeca, each actor has only one thread of execution and its local state is encapsulated from other actors. This greatly simplifies the interactions between actors. But having both continuous and discrete behaviors in one actor, can be considered as having multiple threads of execution in the actor. Since these threads share the same variables, this approach is inconsistent with Rebeca and can surprise the modeler. A simple example to highlight this issue is to consider the following code segment of a message server:
\begin{lstlisting}[xleftmargin=.4\textwidth,language=HRebeca,numbers=none,basicstyle=\footnotesize, frame=none]
		a = k;
		delay(2);
		b = a + c;
\end{lstlisting}
The constant $k$ is assigned to the variable $a$. The delay statement is used to abstractly model the computation time of complex computations. After the specified delay, the value of variable $a$ is used to update the variable $b$. Assume the rebec has a continuous behavior and during the execution of the delay statement, the continuous behavior is finished and changes the value of variable $a$ in its actions. This affects the value of the variable $b$ when the delay statement is over and can lead to a faulty behavior. The separation of software and physical actors, solves this issue. Note that the delay statement is not allowed in the physical actors.

\section{Conclusion and Future Work}\label{Sec:Concolusion}
%Cyber-physical systems are hybrid systems with tight interactions between software and physical processes. There are many applications of CPSs where safety and reliability is central to the design and development of the system. For example, automotive systems are one of the areas where CPSs are becoming more common. Due to the heterogeneity and complexity of these system, new challenges are being faced in the design and development of CPSs.

In this paper we presented an extended actor model for modeling hybrid systems and CPSs, where both continuous and discrete processes can be defined. In this actor model, two kinds of actors are defined: software actors and physical actors. The software actors contain the software behaviors of the model and similarly, physical actors contain the physical behaviors. We also introduced a network entity to the actor model, for modeling the behavior of the network of the model. We implemented this extended actor model in Hybrid Rebeca language. This language is an extension of Timed Rebeca language and allows defining classes to make models modular and reusable. The semantics of the language is defined based on hybrid automata, to allow for formal verification of the models. Since our focus was automotive domain, CAN network is modeled in this version of the language. To show the applicability of our language, we modeled and analyzed a Brake-by-Wire system. For the verification of the model, three safety properties were considered. We used SpaceEx framework to verify these properties. It was shown that for some properties new entities were needed to make the verification feasible. 

%In this work, we addressed the challenges in extending actor model with continuous behaviors. But, some improvements are needed to make Hybrid Rebeca applicable for modeling and verification of CPSs. 
%As it was shown in our case study, the size of the resulting hybrid automaton can be too large for many verification tools. This is due to  our semantics being fine-grained. For the verification of our case study, we abstracted away many instantaneous behaviors and reduced the size of the hybrid automaton drastically. A formal reduction for our language must be defined, to make verification of the models feasible for current tools.

%The semantics of Hybrid Rebeca is defined based on hybrid automata. This leads to two main issues. One issue is that the continuous equations are not solved in semantics level, so all possible permutations for termination of concurrent continuous behaviors are considered and many of these permutations can be unreachable when these equations are solved. The second issue is related to the dynamic variables (e.g. event timers and continuous message parameters) of the model. In hybrid automata these variables must be defined statically so variable pools with defined size are used in our current approach. The size of the pools must be manually defined by the modeler based on the behavior of the model. One possible solution for these two issues is using a lower level model like timed labeled transition system to define the semantics of Hybrid Rebeca. Of course new tools and theories may be needed for verification of these models.

Since we focused on the automotive domain, only CAN network was defined in our current version of the language. Other network models are needed for different applications of CPSs. Instead of defining multiple network models, it's also possible to allow for user-defined network models. To this aim, a set of basic functionalities must be defined to enable defining most network models. Also, defining multiple instances of a network model (e.g. multiple CAN networks) may be needed in some systems. Like network models, it's possible to allow for defining new internal message schedulers for the rebecs, since the FIFO scheduler can be inadequate for some systems.
\section*{Acknowledgments}
We would like to
thank Edward Lee for his supports and patient guidance on modeling and analyzing CPSs, Tom Henzinger for his fruitful discussion on the extended actor model, and MohammadReza Mousavi and Ehsan Khamespanah for their useful contributions.

\bibliographystyle{splncs04}
\bibliography{references}

\end{document}